\documentclass[
 reprint,
 aps,
pra, longbibliography, superscriptaddress,
nobibnotes
]{revtex4-1}
\expandafter\let\csname equation*\endcsname\relax
\expandafter\let\csname endequation*\endcsname\relax
\usepackage{amsmath, amsthm, amssymb}
\usepackage{eqnarray,amsmath,mathtools,nccmath,amssymb}
\usepackage{resizegather}

\usepackage[usenames,dvipsnames]{xcolor}
\usepackage[english]{babel}
\usepackage{lmodern,microtype,ragged2e,comment}
\usepackage{appendix,hyphenat,mathtools,dsfont,bm,bbm}
\usepackage[makeroom]{cancel}
\usepackage{graphicx}
\usepackage{soul}
\usepackage[normalem]{ulem}

\definecolor{mygrey}{gray}{0.35}
\definecolor{myblue}{rgb}{0.2,0.2,0.8}
\definecolor{myzard}{cmyk}{0,0,0.05,0}
\definecolor{mywhite}{rgb}{1,1,1}
\definecolor{myred}{rgb}{0.9,0.1,0.}
\definecolor{dgreen}{rgb}{0.0, 0.5, 0.0}
\definecolor{mypink}{rgb}{1.0, 0.01, 0.24}
\usepackage[colorlinks=true,citecolor=myblue,linkcolor=myblue,urlcolor=myblue]{hyperref}
\usepackage{cleveref}

\DeclareGraphicsExtensions{.png,.pdf,.eps,.jpg}
\graphicspath{ {./figs/} }

\newcommand{\bra}[1]{\langle #1|}
\newcommand{\ket}[1]{|#1\rangle}
\newcommand{\Exp}[1]{\langle#1\rangle}
\newcommand{\proj}[1]{\ket{#1}\bra{#1}}

\newcommand{\braket}[2]{\left\langle #1\lvert#2\right\rangle}
\newcommand{\abs}[1]{\left\lvert #1\right\rvert}

\newcommand{\bematrix}{\left(\begin{matrix}}
\newcommand{\ematrix}{\end{matrix}\right)}

\def\one{{\mbox{$1 \hspace{-1.0mm}  {\bf l}$}}}						
\DeclareMathOperator{\sinc}{sinc}
\DeclareMathOperator{\csch}{csch}

\def\ii{\mathrm{i}}
\def\tr{\mathrm{Tr}}
\def\d{\mathrm{d}}

\def\paulivec{\bm \sigma}
\def\sone{\sigma_1}											
\def\stwo{\sigma_2}
\def\sthree{\sigma_3}
\def\sx{\sigma_x}

\def\cB{\mathcal B}

\def\cF{\mathcal F}

\def\cH{\mathcal H}

\def\cL{\mathcal L}

\def\cN{\mathcal N}
\def\cO{\mathcal O}

\def\cR{\mathcal R}

\newcommand\defn[1]{\textsl{#1}}

\newcommand{\eqnref}[1]{Equation~(\ref{#1})}

\begin{document}

\title{Discrimination and estimation of incoherent sources under misalignment}

\author{J. O. de Almeida}
\affiliation{ICFO-Institut de Ciencies Fotoniques, The Barcelona Institute of Science and Technology, Av. Carl Friedrich Gauss 3, 08860 Castelldefels (Barcelona), Spain}
\email{jessica.almeida@icfo.eu}
\author{J. Ko\l ody\'nski} 
\affiliation{Centre for Quantum Optical Technologies, Centre of New Technologies, University of Warsaw, Banacha 2c, 02-097 Warsaw, Poland}
\author{C. Hirche}
\affiliation{QMATH, Department of Mathematical Sciences, University of Copenhagen, Universitetsparken 5, 2100 Copenhagen, Denmark}
\author{M. Lewenstein}
\affiliation{ICFO-Institut de Ciencies Fotoniques, The Barcelona Institute of Science and Technology, Av. Carl Friedrich Gauss 3, 08860 Castelldefels (Barcelona), Spain}
\affiliation{ICREA-Instituci\'o Catalana de Recerca i Estudis Avan\c cats, Lluis Companys 23, 08010 Barcelona, Spain}
\author{M. Skotiniotis}
\affiliation{F{\'i}sica Te\`{o}rica: Informaci{\'o} i Fen\`{o}mens Qu\`{a}ntics, Departament de F{\'i}sica, Universitat Aut\`{o}noma de Barcelona, 08193 Bellaterra (Barcelona), Spain}


\begin{abstract}
Spatially resolving two incoherent point sources whose separation is well below the diffraction limit dictated by classical optics has 
recently been shown possible using techniques that decompose the incoming radiation into orthogonal transverse modes. Such a 
demultiplexing procedure, however, must be perfectly calibrated to the transverse profile of the incoming light as any 
misalignment of the modes effectively restores the diffraction limit for small source separations. We study by how much can one 
mitigate such an effect at the level of measurement which, after being imperfectly demultiplexed due to inevitable misalignment, 
may still be partially corrected by linearly transforming the relevant dominating transverse modes. We consider two 
complementary tasks:~the estimation of the separation between the two sources and the discrimination between one and two 
incoherent point sources. We show that, although one cannot fully restore super-resolving powers even when the value of the 
misalignment is perfectly known its negative impact on the ultimate sensitivity can be significantly reduced.  In the case of 
estimation we analytically determine the exact relation between the minimal resolvable separation as a function of misalignment 
whereas for discrimination we analytically determine the relation between misalignment and the probability of error, as well as 
numerically determine how the latter scales in the limit of long interrogation times.
\end{abstract}


\maketitle

\section{\label{sec:Intro}Introduction}
Quantum theory has, over the years, exhibited an innate ability to surpass the limitations in performance set by classical devices 
in a variety of tasks~\cite{Gisin2007,Smith2010,Degen2017,Pirandola2018} arguably none more so than in the field of statistical inference and 
decision theory.  There the use of distinctive quantum features, such as coherence and entanglement, allows for the existence of 
ultra-precise measurements~\cite{Degen2017} that greatly enhance the performance in a variety of sensing tasks---ghost 
imaging~\cite{Lemos2014} and quantum illumination~\cite{Lloyd2008} to name but a few---that are impossible to achieve by even 
the best classical means.     
 
One such success of the quantum mechanical formalism concerns the spatial resolution of imaging devices.  For over a century it 
was believed that two sources of incoherent light can barely be resolved if ``the maximum of one is over the minimum of the 
other''~\cite{Rayleigh:1879}; any closer than this and conventional classical imaging techniques cannot resolve the two incoherent 
sources, even if an asymptotically large number of photons are detected. Despite several 
efforts~\cite{Acuna2002,Shahram2004,Ram2006,Shahram2006} this limitation of optical imaging systems---known as the 
\emph{diffraction} or \emph{Rayleigh} limit~\cite{Rayleigh:1879}---seemed insurmountable until a proper quantum mechanical 
treatment of the problem revealed that, just like many other classically derived limitations, it too can be 
overcome~\cite{Tsang2016}.  Rather than imaging directly the incoming radiation it was proven that a simple linear-optical 
preprocessing of the spatial profile of the electromagnetic field into a predefined set of spatially orthogonal modes, e.g., the 
Hermite-Gauss modes in case of Gaussian apertures~\cite{Svelto2010}, followed by photon detection over sufficiently long 
integration time is capable of resolving two incoherent point sources at arbitrary separation. The reason for this drastic 
improvement is intuitive: spatially orthogonal modes of light provide information about spatial correlations of the incoming 
photons, whereas direct imaging does not. 

The technique of decomposing, or demultiplexing, the optical field into spatially orthogonal modes followed by photon counting 
has gained increased attention with rapid theoretical and experimental developments (see~\cite{Tsang2019} for a recent review). 
Its performance has been proven not only in complex \emph{estimation} tasks, such as resolving multiple 
sources~\cite{Tsang2017,Tsang2019b,Zhou2019,Bisketzi2019,Sidhu2017,sidhu2018}, sources of unequal 
brightness~\cite{Rehacek:2017,Rehacek:2018}, sources emitting coherent~\cite{Tsang2019c} or non-classical~\cite{Lupo2016} 
light, as well as sources localised arbitrarily in space~\cite{Prasad2019}, but also for the closely related problem of 
\emph{discrimination} beyond the diffraction limit~\cite{Lu2018}.

\begin{figure*}[t!]  
\centering
   \includegraphics[keepaspectratio, width=0.95\textwidth]{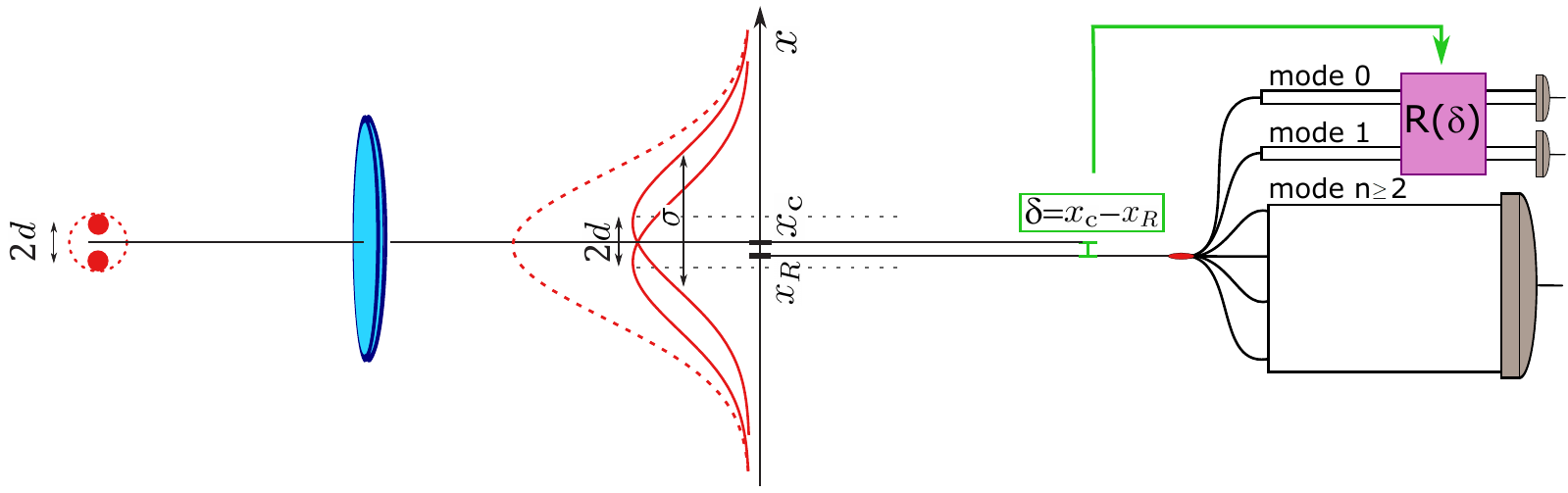}
   \caption{\emph{Super-resolving the separation between incoherent sources under the misalignment of the imaging apparatus.} Two incoherent point-like sources of light are imaged with an optical system exhibiting a Gaussian point spread function of width $\sigma$ in a way that their separation, $2d$, can be most accurately resolved. For this to be possible beyond the diffraction limit, a spatial mode demultiplexing technique is employed---which 
   ideally allows the incoming light to be decomposed into orthogonal transverse modes, whose photon-occupation is subsequently measured. In this work, we study the ultimate limits on the resolution in the presence of misalignment of the imaging system, $\delta = x_{\mathrm{c}}-x_R\ll\sigma$, by applying appropriate linear optical post processing operations $R(\delta)$ is applied on the two dominant modes of the demultiplexing measurement.} 
   \label{fig:ROTADE}
\end{figure*}

Moreover, the robustness to imperfections of the proposed schemes has recently been an object of intensive 
research~\cite{Len2019,Lupo2019,gessner2020}, largely motivated by the challenges imposed by up-to-date experimental 
demonstrations~\cite{paur_achieving_2016,yang_far-field_2016,Tham2017,donohue_quantum-limited_2018,Parniak2018,zhou_quantum-limited_2019,paur_tempering_2018}. 
An important obstacle pointed out in the original paper of \emph{Tsang et al.}~\cite{Tsang2016} is the crucial assumption that the 
centroid ($x_{\mathrm{c}}$)---the midpoint between the two light sources whose separation, $2 d$, is to be resolved---is perfectly aligned 
($x_{\mathrm{c}}=x_R$) with the detector position ($x_R$)---where the spatial transverse modes are demultiplexed. In the presence 
of any misalignment, $\delta=x_{\mathrm{c}}-x_R$ (see Fig.~\ref{fig:ROTADE}), the Fisher information, $F$, that quantifies the 
ultimate resolution no longer approaches a constant with vanishing separation but rather behaves as $F\sim(d/\delta)^2\to0$ for 
$d\to 0$~\cite{Tsang2016}.

Although the location of the centroid can be estimated via direct imaging, this requires sacrificing an, in principle large amount, of 
photons in order to do so. A way of leveraging the use of photons between direct imaging and spatial mode 
demultiplexing, using information of the former to better position the latter, has been recently proposed~\cite{Grace2019}.  Here,
in contrast, we take misalignment as experimental fact and investigate the theoretical limits imposed by any misalignment in both 
the canonical \emph{estimation} task~\cite{Tsang2016,Len2019,Lupo2019,gessner2020,paur_achieving_2016,yang_far-field_2016,Tham2017,donohue_quantum-limited_2018,Parniak2018,zhou_quantum-limited_2019,paur_tempering_2018,Grace2019} as well as the complementary task of 
\emph{hypothesis testing}~\cite{Krovi2016,Lu2018}.

Specifically, we ask the following questions;  how do estimation precision and the minimal resolvable distance scale as a function 
of misalignment in the canonical estimation task and, how does the probability of error---both in single observation, as well as 
asymptotically---depend on misalignment in a typical hypothesis testing task.  We keep our treatment as general as possible by 
considering the use of passive linear optics \emph{after} the demultiplexing stage (see Fig.~\ref{fig:ROTADE}). We show that 
even if the value of misalignment $\delta$ is perfectly known such linear optical post processing is not capable of restoring the super-resolution.  We demonstrate that optical post-processing of the two most dominants modes, 
tailored to the misalignment $\delta$, improves estimation precision over the ``raw" demultiplexed measurements  from 
$F\propto1/\delta^2$ to $F\propto1/\delta^6$ as $\delta\approx 0$ and the minimal resolvable distance from 
$\epsilon_{\min}\propto\delta^{\frac{1}{2}}$ to $\epsilon_{\min}\propto\delta^{\frac{3}{2}}$ for both Gaussian and 
Sinc point-spread functions. Moreover, for single-shot ($n=1$) discrimination the 
probability of erroneously interpreting a single source for a double one scales as $P_\textrm{err}\propto \delta^6$ compared to 
$P_\textrm{err}\propto \delta^2$ for the misaligned demultiplexed measurements for small $\delta\approx0$. Furthermore, we 
also show that linear post-processing also helps in the asymptotic $n\to\infty$ regime, where we observe an enhancement of up 
to $12\%$ in the exponential decay of the total probability of error.

One may object that if the misalignment is known, then super-resolution can easily be restored simply by adjusting the 
demultiplexing device to the right position. We stress, however, that in our work this assumption is made in order to explore the 
theoretical limits on super-resolution allowed by the most general post-processing \emph{after} an imperfect demultiplexing 
procedure.  From this point of view, our results directly link pertinent quantities, the quantum Fisher information, minimal 
resolvable distance, and probability of error, to the value of the misalignment.  From the practical point of view even if the 
position of the centroid is perfectly known, fabrication of the demultiplexing measurements, or adjustment of their position via 
mechanical servo mechanisms entails an inherent uncertainty~\cite{Morizur:10} thus not allowing for perfect alignment.  
In such instances our theoretical analysis offers insight to additional strategies, which may be implemented efficiently by 
modulating the phase of integrated waveguides~\cite{wang2019integrated,he2019high}, that could be used in conjunction to 
current experimental proposals~\cite{Grace2019}. 

The article is structured as follows.  Firstly, we review the necessary mathematical background for both classical and quantum 
mechanical image resolution in Section~\ref{sec:Background}. In Section~\ref{sec:parest} we study the effects of misalignment for 
the problem of estimating the separation between two incoherent point sources, whilst Section~\ref{sec:hypothesistesting} deals 
with the effects of misalignment for the problem of discriminating the one versus two sources hypothesis.  
Section~\ref{sec:conclusion} summarizes our work and discusses possible future directions of investigation.

\section{\label{sec:Background} Diffraction Limited Optical Imaging}
 
We begin by reviewing the mathematical treatment of optical imaging devices.  In Section~\ref{sec:Rayleigh} we review imaging 
in classical optics paying particular attention on how the diffraction limit comes about in these set-ups.  In Section~\ref{sec:model} 
we give a formal quantum mechanical description of the \emph{point spread function} (PSF) of an optical imaging system.  We 
shall restrict our attention particularly to one-dimensional Gaussian and Sinc PSFs but the analysis easily extends to other PSFs 
and to higher dimensions.  We then review a mathematical approximation to the quantum mechanical state of the PSF---the qubit 
model of Chrostowski {\it et al.}~\cite{Chrostowski2017}.  The latter will be used to explore how misalignment of the optical 
imaging system affects its performance, as well as to propose alternative measurement schemes that compensate for 
misalignment. 

\subsection{\label{sec:Rayleigh} Classical theory of diffraction limited optical imaging}

To image light sources that are far away requires specific lens and aperture systems that allow to process the spatial distribution 
of the emitters. Assuming the paraxial approximation holds diffraction effects cause variations in radiation 
intensity at the image plane---the familiar bright and dark fringes in imaging stars, or diffraction gratings. Consequently, the 
minimum angular distance between two or more emitters that allows their distinction---the \emph{angular resolution} of the 
imaging device---is fundamentally limited due to diffraction. Lord Rayleigh was the first to obtain a heuristic rule for the angular 
resolution of any imaging device~\cite{Rayleigh:1879}:  two point sources can barely be resolved so 
long as the central maximum in intensity of one source lies on top of the first minimum in intensity of the second in the image 
plane.  This rule of thumb is colloquially known as Rayleigh's curse or \emph{diffraction} limit in optical imaging.

For the simplest optical imaging device consisting of a single slit of width $D$ the diffraction limit can be deduced by simple 
geometrical optics, and corresponds to the angular distance, $\phi$, between the central intensity maximum and first minimum 
which is given by
\begin{equation}
\phi\approx\frac{\lambda}{D}
\label{eq:Rayleigh_geom}
\end{equation}  
where $\lambda$ is the wavelength of the incoming radiation, and the approximation sign is due to the paraxial approximation.  

More formally, the diffraction limit can be obtained by making use of the Fresnel-Kirchoff formula which describes the amplitude 
of the disturbance in a given direction, $\phi$, from the optical axis due to the aperture of the imaging 
system~\cite{Fowles1989}.  For a one-dimensional aperture whose profile is given by $f(y)$, the Fresnel-Kirchoff formula reads
\begin{equation}
\Psi(\phi)=\frac{1}{\sqrt{2\pi}}\int_{-\infty}^{\infty} f(y) e^{\ii k\, y\sin \phi}\, \d y,
\label{eq:FK}
\end{equation}
where $k=2\pi/\lambda$ is the wavenumber. The intensity distribution, also known as the objects \emph{point spread function} (PSF), at angular separation $\phi$ is given by $\abs{\Psi(\phi)}^2$, and we have implicitly assumed the intensity is normalized $\int \abs{\Psi(\phi)}^2\d \phi=1$.
  Eq.~\eqref{eq:FK} is the familiar statement that the PSF at the image plane of an image system is the Fourier transform of the systems aperture.  The case of the single slit of width $D$ corresponds 
to $f(y)=\mathrm{rect}\left(-\frac{D}{2},\frac{D}{2}\right)$ and gives rise to the Sinc PSF
\begin{equation}
\Psi(\phi)\propto\mathrm{sinc}\left(\frac{Dk\sin\phi}{2}\right),
\label{eq:singleslitPSF}
\end{equation}
here the sinc function is defined as $\mathrm{sinc}(x)=\frac{\sin(x)}{x}$. The first minimum of the sinc function occurs at $\frac{Dk\sin\phi}{2}=\pi$, i.e., $\phi\approx\frac{\lambda}{D}$ which is the 
 result obtained using geometric optics. For a circular aperture $f(y)=\sqrt{D^2-4y^2}$, where $D$ is the diameter of the 
aperture, the corresponding PSF reads 
\begin{equation}
\Psi(\phi)\propto \frac{J_1\left(\frac{Dk}{2}\sin\phi\right)}{\frac{Dk}{2}\sin\phi},
\label{eq:circular}
\end{equation}
where $J_1(z)$ is the Bessel function of the first kind. The first minimum of the latter occurs when 
$\frac{\pi D\sin\phi}{\lambda}=3.8317$, which sets the angular resolution to $\phi\approx\frac{1.22\lambda}{D}$.

One can, in principle, shape the PSF of an imaging system to any desired function using apodization that suppresses the higher 
order intensity maxima of the diffraction pattern~\cite{Debes2004}.  Such techniques can be used to turn the Bessel function 
PSF of the circular aperture to a Gaussian one. As such techniques do not alter the shape of the aperture, the diffraction limit 
above still holds.
 
\subsection{\label{sec:model}Quantum description of two incoherent point sources}

Consider two incoherent point sources (e.g., stars or bacteria fluorescing) emitting monochromatic light.  
We shall assume that the sources are weak, meaning that the average number of photons detected by our 
imaging device is much smaller than one.  Quantum mechanically we may represent the state of the incoming 
radiation by the density operator~\cite{Tsang2016}:
\begin{equation}
\sigma^{(i)}\approx (1-\varepsilon)\,\proj{0}+\varepsilon\,\rho^{(i)}+\cO(\varepsilon^2),
\label{eq:weak_source_approximation}
\end{equation}
where $\varepsilon\ll 1$, $\proj{0}$ corresponds to the vacuum state, and $\rho^{(i)}\in\cB(\cH_1),\, i\in(1,2)$ is a one-photon state 
with the superscript index labelling the case where the photon is due to one or two point sources. 

As the vacuum offers no information about the nature of the emitting source our only information comes from the single photon
events, accumulated over sufficiently long time, at the image plane of our instrument.  Assuming the latter to be  
one-dimensional we define the image plane position eigenkets $\ket{x}=a^\dagger(x)\ket{0}$, where $a^\dagger(x),\, a(x)$ 
are the creation and annihilation operators satisfying $[a(x),a^\dagger(y)]=\delta(x-y)$~\cite{Yuen1978}.  The wave function of a 
single photon can now be expanded in terms of the position basis of the image plane as
\begin{equation}
\ket{\Psi(z)}=\int_{-\infty}^\infty \d x\, \Psi(x-z)\ket{x},
\label{eq:PSF_ampl}
\end{equation}
where $\abs{\Psi(x-z)}^2=\abs{\braket{x}{\Psi(z)}}^2$ denotes the probability of detecting a photon at position $x$ in the image 
plane---the objects PSF. $\Psi(x-z)$ is equivalent to $\Psi(\phi)$ of Eq.~(\ref{eq:FK}) in the far field regime in which $\phi\approx0$ and, hence, $\sin(\phi)\approx\phi$.

For a Gaussian or square aperture the PSF of a single incoherent point source is the corresponding Fourier 
transform~\cite{Svelto2010,Kerviche2017},
\begin{subequations}
\begin{align}
\abs{\Psi(x-z)}^2&=\frac{1}{\sqrt{2\pi\sigma^{2}}}\exp{\left(-\frac{(x-z)^{2}}{2\sigma^{2}}\right)},\label{eq:single_source_PSFa}\\
\abs{\Psi(x-z)}^2&=\frac{1}{\sigma}\sinc^2\left(\pi\frac{x-z}{\sigma }\right)\label{eq:single_source_PSFb},
\end{align}
\end{subequations}
respectively, where $z$ is the mean of the PSF and $\sigma^2$ the corresponding variance (both fully characterised by the 
imaging system), and we have introduced appropriate normalisation factors ensuring that $\int \abs{\Psi(x-z)}^2\d x=1$.  
The variance is taken to be $\sigma^2\approx\frac{\lambda^2}{D^2}$, where $D$ is the diameter (length) of the Gaussian 
(square) apertures, respectively Eqs.~(\ref{eq:single_source_PSFa}) and (\ref{eq:single_source_PSFb}). Note Eq.~(\ref{eq:single_source_PSFb}) is equivalent to Eq.~(\ref{eq:singleslitPSF}) in the far field regime, where $x-z=\sin\phi\approx\phi$. 

The state of a single photon emanating from a single point source whose PSF is centred around $z=x_0$ is then described by 
the state
\begin{equation}
\rho^{(1)}=\proj{\Psi(x_0)},
\label{eq:single_source}
\end{equation}
whereas for a photon coming from two incoherent point sources with relative intensities $w$ and $1-w$, whose PSF's are 
centred around $x_1$ and $x_2$, is described by the density matrix 
\begin{equation}
\rho^{(2)}=w\proj{\Psi(x_1)}+(1-w)\proj{\Psi(x_2)}.
\label{eq:two_sources}
\end{equation}  
For the case of two incoherent point sources it is convenient to define the \emph{centroid}
\begin{equation}
x_{\mathrm{c}}:=w\, x_1+(1-w)x_2,
\label{eq:xc}
\end{equation}
and \emph{separations}
\begin{equation}
d_i:=\abs{x_i-x_{\mathrm{c}}}.
\label{eq:d}
\end{equation}
For two sources of equal intensity---the case shown in Fig.\ref{fig:ROTADE} that we will focus on hereafter---the centroid and 
separations read  $x_{\mathrm{c}}=\frac{x_1+x_2}{2}$, and $d_1=d_2=d=\abs{\frac{x_2-x_1}{2}}$ respectively.  It is also often 
assumed that the mean of the PSF for a single incoherent point source coincides with the centroid of two point sources, i.e., 
$x_{\mathrm{c}}=x_0$.

\begin{figure}[t!]  
\centering
   \includegraphics[keepaspectratio, width=0.45\textwidth]{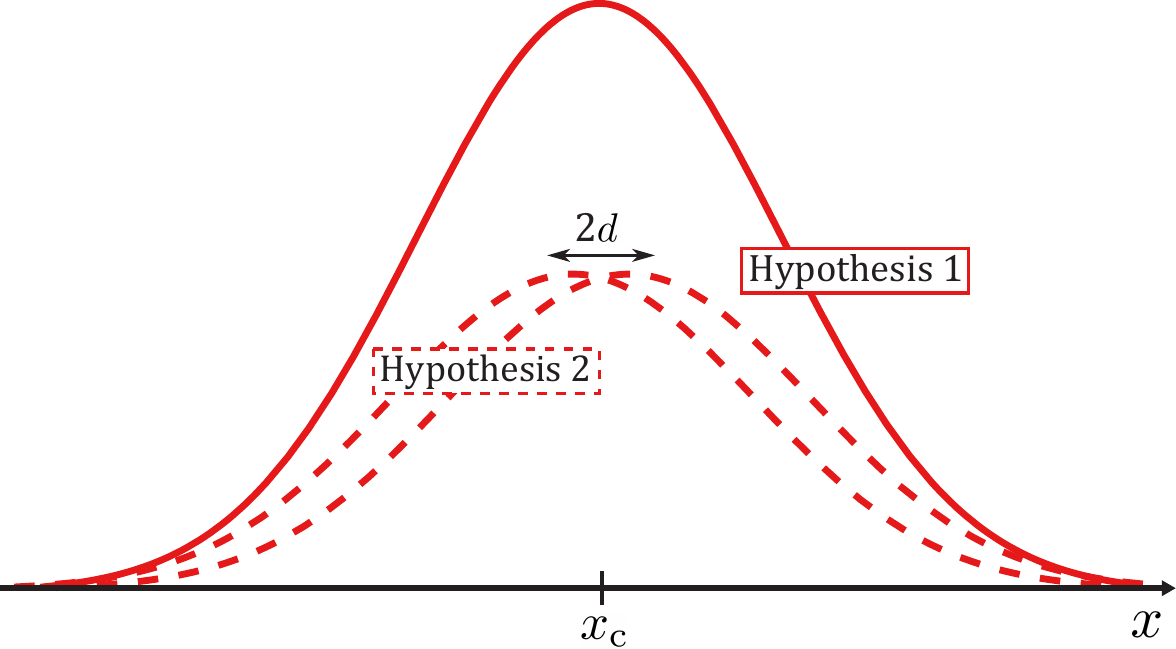}
   \caption{Intensity distribution at the image plane arising from two nearly coinciding incoherent light sources ($\sigma\gg d$) 
   given an imaging system that exhibits a Gaussian point spread function (PSF). Due to the large overlap between the two PSFs, 
   direct imaging does not allow to accurately \emph{estimate} the positions of each source, and is incapable of 
   \emph{discriminating} whether the image is the result of one or two sources---Hypothesis 1 and Hypothesis 2 respectively.}
   \label{fig:Rayleigh's_curse}
\end{figure}

If $\sigma\ll d$, then both the centroid and separation can be effectively estimated via conventional means, 
specifically by direct imaging~\cite{Rayleigh:1880}. However, for $\sigma\gg d$---the relevant regime we explicitly depict in 
Fig.\ref{fig:Rayleigh's_curse}---the diffraction limit implies that the two sources cannot be resolved even if we observe 
asymptotically many photons~\cite{Rayleigh:1879}.

In order to overcome the diffraction limit Tsang {\it et al.}~\cite{Tsang2016} proposed to abandon direct imaging and count instead 
the number of photons in distinct spatial modes of light. In particular, when dealing with Gaussian PSFs the spatial modes can be 
interpreted as the energy eigenstates of the quantum mechanical harmonic oscillator, i.e., the Hermite-Gauss (HG) modes: 
\begin{equation}
\ket{\Phi_n(x_R)}=\frac{1}{\sqrt{2^n\, n!}}\frac{1}{\sqrt[4]{2\pi\, \sigma^{2}}}\times\int_{-\infty}^{\infty}e^{-\frac{(x-x_R)^2}
{4\sigma^{2}}}\, H_n\!\left(\frac{x-x_R}{\sqrt{2}\sigma}\right)\, \ket{x}\,\d x,
\label{eq:HOeigenstates}
\end{equation}
where
\begin{equation}
H_n(\alpha):=(-1)^n\, e^{\alpha^2}\frac{\d^n}{\d\alpha^n}\,e^{-\alpha^2}
\label{eq:Hermitepolynom}
\end{equation}
are the Hermite polynomials, and $x_R$ is the \emph{reference position} of the spatial modes. This measurement can be 
implemented with the help of linear optical pre-processing of the incoming radiation, followed by photon-number resolving 
detectors and can resolve two point sources no matter how close their PSFs are on the image plane so long as
$x_R=x_{\mathrm{c}}$, i.e.,~the position of their centroid is known exactly. The latter can be estimated via direct imaging; its 
precision varies inversely proportionally with the square root of the measurement integration time~\cite{Tsang2016}.  Moreover, 
even if the centroid  is known sufficiently well, there is still the issue of perfectly aligning the measurement device for spatial mode 
demultiplexing, or SPADE for short, i.e., setting $x_R=x_{\mathrm{c}}$. A proposal on how best to accomplish this using a finite 
number of observations was proposed recently by Grace {\it et al.}~\cite{Grace2019}.  There the authors proposed to combine direct imaging and SPADE techniques, in a two-stage procedure; the former uses part of the incoming radiation to adjust the exact position of the latter via a servo feedback mechanism in order to gradually reduce the \emph{misalignment}, $\delta:=x_{\mathrm{c}}-x_R$ marked explicitly in Fig.~\ref{fig:ROTADE}, which is \emph{a priori} not known.   

In this work we take misalignment of the demultiplexing device as fact and determine the theoretical limits imposed by such
misalignment in both \emph{estimation} and \emph{discrimination} tasks (see Fig.~\ref{fig:Rayleigh's_curse}).  To do so we allow ourselves the freedom of performing arbitrary, linear-optical  post processing of the demultiplexed radiation. Specifically, we shall analytically determine how such linear-optical post processing, based  on complete knowledge of the value of the misalignment, affects  estimation precision, as well as the minimal resolvable distance.
To do so we shall make use of an approximation of the state of the incoming radiation known as the qubit model~\cite{Chrostowski2017}, which we now review. 

\subsection{\label{sec:qbitmodel} The qubit model for two incoherent point sources.}

The qubit model is an approximation of the PSF in the presence of misalignment~\cite{Chrostowski2017}.  The latter can be 
understood as performing the projective measurement of Eq.~\eqref{eq:HOeigenstates} about some reference position 
$x_R\neq x_i$ for $i\in(0,1,2)$. Assuming that this misalignment is small, i.e., $x_R\approx x_i$, we can Taylor expand the 
probability amplitudes of each source, $\Psi(x-x_i), \, i\in\{1,2\}$, about $x_R$ as follows:
\begin{equation}
\begin{split}
\ket{\Psi(x_i)}\approx&\int_{-\infty}^{\infty} \d x\, \Psi(x-x_R)\,\ket{x}+(x_i-x_R)\int_{-\infty}^{\infty} \d x\,\frac{\d \Psi(x-x_i)}{\d x_i}\bigg\rvert_{x_{i}=x_{R}}\ket{x}\\
=: &\; \ket{0}-(x_i-x_R)\sqrt{\cN}\ket{1},
\end{split}
\label{eq:Taylor_approx}
\end{equation}
and identify a qubit subspace with $\ket{0}:=\ket{\Psi(x_R)}$ and
\begin{equation}
\ket{1}:=\frac{-1}{\sqrt{\cN}}\int_{-\infty}^{\infty} \d x\,\frac{\d \Psi(x-x_i)}{\d x_i}\bigg\rvert_{x_{i}=x_{R}}\ket{x}
\label{eq:onestate}
\end{equation}
an orthonormal basis. Here, $\cN$ is an appropriate normalisation factor which for the Gaussian and Sinc PSFs 
reads 
\begin{equation}
\begin{split}
\cN_{\mathrm{G}}&=\frac{1}{4\sigma^2},\qquad
\cN_{\mathrm{S}}=\frac{\pi^2}{3\sigma^2},
\end{split}
\label{eq:norms}
\end{equation}
respectively.

The state of the incoming radiation can now be described, to a very good approximation, by the following qubit density operators, 
for one and two sources, respectively:
\begin{align}  \nonumber
\rho^{(1)}&\approx\frac{1}{1+(\sigma\theta)^2\cN}\bematrix
1  & -\sigma\theta\sqrt{\cN}\\
 -\sigma\theta\sqrt{\cN} &(\sigma\theta)^2\cN
\ematrix\\
\rho^{(2)}&\approx\frac{1}{1+\sigma^2(\theta^2+\epsilon^2)\cN}\bematrix
1  & -\sigma\theta\sqrt{\cN}\\
- \sigma\theta\sqrt{\cN} &\sigma^2(\theta^2+\epsilon^2)\cN
\ematrix,
\label{eq:qubitmodel}
\end{align} 
where we now introduced dimensionless parameters for \emph{misalignment} and \emph{separation}:
\begin{equation}
\theta:=\frac{\delta}{\sigma}=\frac{x_{\mathrm{c}}-x_R}{\sigma}
\qquad\text{and}\qquad
\epsilon:=\frac{d}{\sigma},
\label{eq:parC}
\end{equation}
respectively. The qubit model allows us to visualise the effects of misalignment on a given PSF in terms of the Bloch 
representation of qubit density matrices, i.e.,
\begin{equation}
\rho:=\frac{\one+\bm{{\bf r}}\cdot\paulivec}{2},
\label{eq:Blochrep}
\end{equation}
where $\bm{{\bf r}}\in\cR_3$, has elements $\mathrm{r}_i=\tr(\sigma_i\rho)$ and $\paulivec:=(\sone,\stwo,\sthree)^T$ is the 
vector of Pauli matrices $\sigma_i$.  For the Gaussian and Sinc PSFs the corresponding Bloch vectors read
\begin{align}\nonumber
{\bf r}^{(1)}_\mathrm{G}&=\frac{1}{1+\frac{\theta^2}{4}}\bematrix
-\theta\\
0\\
1-\frac{\theta^2}{4}
\ematrix,\;
{\bf r}^{(2)}_\mathrm{G}=\frac{1}{1+\frac{\theta^2+\epsilon^2}{4}}\bematrix
-\theta\\
0\\
1-\frac{\theta^2+\epsilon^2}{4}\\
\ematrix\\
{\bf r}^{(1)}_\mathrm{S}&\approx\frac{1}{1+\frac{\theta^2}{3}}\bematrix
-\frac{2\theta}{\sqrt{3}}\\
0\\
1-\frac{\theta^2}{3}
\ematrix,\;
{\bf r}^{(2)}_\mathrm{S}\approx\frac{1}{1+\frac{\theta^2+\epsilon^2}{3}}\bematrix
-\frac{2\theta}{\sqrt{3}}\\
0\\
1-\frac{\theta^2+\epsilon^2}{3}
\ematrix,
\label{eq:blochvecGaus}
\end{align}
respectively. Using the approximations
\begin{equation}
\begin{split}
\frac{1}{1+x^2}&\approx 1-x^2\\
1-\frac{(\theta^2+\epsilon^2)}{2}&\approx(1-\frac{\theta^2}{2})(1-\frac{\epsilon^2}{2})\approx \cos\theta\left(1-\frac{\epsilon^2}{2}\right),
\end{split}
\label{eq:approx}
\end{equation}
and keeping terms up to second  order, $\cO(\theta^i\epsilon^j)$ with $i+j=2$, the Bloch vectors in 
Eq.~\eqref{eq:blochvecGaus} can be further approximated by
\begin{align}\nonumber
{\bf r}^{(1)}_\mathrm{G}&\approx\bematrix
-\sin\theta\\
0\\
\cos\theta
\ematrix,\quad
{\bf r}^{(2)}_\mathrm{G}\approx\left(1-\frac{\epsilon^2}{2}\right)\bematrix
-\sin\theta\\
0\\
\cos\theta
\ematrix\\
{\bf r}^{(1)}_\mathrm{S}&\approx\bematrix
-\sin\frac{2\theta}{\sqrt{3}}\\
0\\
\cos\frac{2\theta}{\sqrt{3}}
\ematrix,\quad
{\bf r}^{(2)}_\mathrm{S}\approx\left(1-\frac{\epsilon^2}{2}\right)\bematrix
-\sin\frac{2\theta}{\sqrt{3}}\\
0\\
\cos\frac{2\theta}{\sqrt{3}}
\ematrix.
\label{eq:blochvecGausAppr}
\end{align}
Consequently the misalignment, $\theta$, can be understood as an infinitesimal rotation about the $y$-axis 
in the Bloch-sphere picture, whereas the separation, $\epsilon$, between the centres of the two incoherent point sources 
affects the purity of the state~\cite{Chrostowski2017}.

Our aim is to use the qubit model to study the effects of misalignment, both in the estimation of the separation between two 
point sources, as well as in the task of discriminating between the single and two source hypotheses. We begin first with 
estimating the separation between two incoherent point sources.

\section{\label{sec:parest}Separation estimation under misalignment}

In this section we review the quantum information tools for multi-parameter estimation, after which we use the qubit model to 
derive the optimal measurement for estimating the separation between two incoherent point sources under misalignment. 

\subsection{\label{sec:CramerRao}Classical and quantum statistical inference}

The task at hand is the estimation of two parameters: the two sources centroid position $x_{\mathrm{c}}$, and their 
separation $d$ from a finite sample of $n$ measurement outcomes $\bm{y}:=(y_1,\ldots,y_n)^T,\, y_i\in\mathbb{R}$, in one 
dimension~\footnote{The results we mention also hold for multiple dimensions.}.  For ease of notation let us 
denote the parameters to be estimated by $\bm{\lambda}:=(\lambda_1,\lambda_2)^T\in\mathbb{R}^{ 2}$.  Then the data 
constitutes a random variable $\bm{y}\in\bm{Y}$ distributed according to $p(\bm{y}|\bm{\lambda})$.

An estimator, $f_i:\bm{Y}\to\mathbb{R}$, is any function that maps every possible measurement record to an estimate 
$\hat{\lambda}_i=f_i(\bm{y})$ of the parameter $\lambda_i$. An estimator is said to be \defn{unbiased} if 
$\Exp{\hat{\lambda}_i}:=\sum_{\bm{y}}p(\bm{y}|\bm{\lambda})\hat{\lambda}_i=\lambda_i$.  Denoting by 
$\hat{\bm{\lambda}}\in\mathbb{R}^{2}$ the two-dimensional vector of estimates of $\bm{\lambda}$, 
the Cram\'{e}r-Rao inequality places a lower bound on the covariance matrix of any unbiased estimator~\cite{Cramer:61} :
\begin{equation}
    \left\langle\left(\hat{\bm{\lambda}}-\bm{\lambda}\right)\cdot\left(\hat{\bm{\lambda}}-\bm{\lambda}\right)^T\right\rangle\geq
    \left(n\,\bm{F}\left(p(\bm{y}|\bm{\lambda})\right)\right)^{-1},
    \label{eq:ClassicalCRbound}
\end{equation}
where $\bm{F}\left(p(\bm{y}|\bm{\lambda})\right)$ is the Fisher information matrix~\cite{Fisher:1922}
\begin{equation}
    F_{ij}(p(\bm{y}|\boldsymbol{\lambda})):=\left\langle\left(\frac{\partial\,\ln p(\bm{y}|\bm{\lambda})}{\partial_{\lambda_i}}\right)\left(\frac{\partial \,\ln p(\bm{y}|\bm{\lambda})}{\partial_{\lambda_{j}}}\right)\right\rangle,
    \label{eq:Classical_Fisher}
\end{equation}
quantifying the amount of information the random variable $\bm{Y}$ carries about the parameters $\bm{\lambda}$.

An estimator is said to be \defn{efficient} if it saturates the inequality of Eq~\eqref{eq:ClassicalCRbound}.
Note that it is possible that no efficient estimator exists if the data sample is finite.  However, for an 
asymptotically large sample size, i.e., $n\to\infty$, it can be shown that the maximum likelihood estimator always 
saturates the Cram\'{er}-Rao bound~\cite{Wilks:62}. 
 
In quantum statistical inference the random variable $\bm{Y}$ and its corresponding probability distribution 
$p(\bm{y}|\bm{\lambda})$ arise from performing a quantum measurement on a quantum system.  Any set of 
\emph{positive operators}, $\{E_{\bm y}\geq 0\, ;\, {\bm y}\in{\bm Y}\}$, satisfying the completeness relation 
$\sum_{\bm y\in Y} E_{\bm y}=\one$ is an admissible measurement, termed a \emph{Positive Operator Valued Measure}, 
or  POVM for short. By virtue of positivity $E_{\bm y}=M^\dagger_{\bm y}\,M_{\bm y}$, where $M_{\bm y}$ constitute one of 
the infinitude of square roots of $E_{\bm y}$.  If $M_{\bm y}=M^\dagger_{\bm y}$ and $M_{\bm y}^2=M_{\bm y}$ then the POVM 
consists of projective operators, and there exists a dynamical variable---energy, position, (angular) momentum, 
{\it etc.}---represented by the Hermitian operator $O$, such that $O=\sum_{\bm y} \mu_{\bm y}\, M_{\bm y}$. 
Given a POVM the conditional probability of obtaining a given measurement record $\bm y$ is given by
\begin{equation}
    p(\bm{y}|\bm{\lambda})=\tr\left(E_{\bm{y}}\rho(\bm{\lambda})\right).
    \label{eq:Bornrule}
\end{equation}
Using the natural Riemannian geometry of the space of bounded, positive linear operators one can define the operator analogue 
of the logarithmic derivative in Eq.~\eqref{eq:Classical_Fisher} for each parameter $\lambda_i$---the \defn{symmetric logarithmic 
derivative} (SLD), $\cL_{\lambda_{i}}$---as the solution to
\begin{equation}
\frac{\partial\,\rho(\bm{\lambda})}{\partial_{\lambda_i}}:=\frac{1}{2}\left(\cL_{\lambda_{i}}\rho(\bm{\lambda})+\rho(\bm{\lambda})
\cL_{\lambda_{i}}\right).
\label{eq:SLD_def}
\end{equation} 

In the eigendecomposition of $\rho(\bm{\lambda)}$, $\{\mu_j,\,\ket{\psi_j}\}$, the SLD operator $\cL_{\lambda_i}$ is 
explicitly given by~\cite{Helstrom1976,Holevo1982} 
\begin{equation}
    \cL_{\lambda_{i}}=2\sum_{\substack{\alpha,\beta\\ \mu_\alpha+\mu_\beta\neq0}}\frac{\bra{\psi_{\alpha}(\bm{\lambda})}
    \partial_{\lambda_{i}}\rho_{\boldsymbol{\lambda}}\ket{\psi_{\beta}(\bm{\lambda})}}{\mu_\alpha(\bm{\lambda})+
    \mu_\beta(\bm{\lambda})}\ket{\psi_{\alpha}(\bm{\lambda})}\bra{\psi_{\beta}(\bm{\lambda})},
    \label{eq:sld}
\end{equation}
and the quantum Fisher information matrix elements read 
\begin{equation}
     \bm{\cF}_{ij}(\rho(\bm{\lambda}))=\frac{1}{2}\tr\left(\rho_{\bm{\lambda}}\left\{\cL_{\lambda_{i}},\cL_{\lambda_{j}}\right\}\right),
     \label{eq:QFIij}
\end{equation}
where $\{A,B\}=AB+BA$.  We thus have the following chain of inequalities for the covariance matrix 
\begin{align}\nonumber
 \left\langle\left(\hat{\bm{\lambda}}-\bm{\lambda}\right)\cdot\left(\hat{\bm{\lambda}}-\bm{\lambda}\right)^T\right\rangle&\geq (n\,
 \bm{F}\left(p(\bm{y}|\bm{\lambda})\right))^{-1}\\
 &\geq (n \, \bm{\cF}\left(\rho(\bm{\lambda})\right))^{-1},
  \label{eq:QuantumCRbound}
\end{align}
the latter inequality commonly referred to as the quantum Cram\'{e}r-Rao bound.

For each single parameter $\lambda_i$ an asymptotically efficient estimator exists and is given by the maximum 
likelihood estimator of the POVM whose elements are the eigenprojectors of the corresponding SLD operator.  
If all these operators commute, i.e., $[\cL_{\lambda_i},\,\cL_{\lambda_j}]=0,\, \forall\, i\neq j$, then the quantum Cramer-Rao 
bound is asymptotically achievable.  Note that commutativity is only a sufficient condition;  a necessary and sufficient 
condition---assuming asymptotically many independent and identically distributed copies ($n\gg1$) of $\rho(\bm{\lambda})$---is 
$\tr\left(\rho(\bm{\lambda})\left[\cL_{\lambda_i},\,\cL_{\lambda_j}\right]\right)=0,\, \forall\, i\neq j$~\cite{Ragy2016}. 
However, note that the POVM that saturates the quantum Cramer-Rao bound in Eq.~\eqref{eq:QuantumCRbound} may, in 
general, correspond to a collective measurement on all the $n\gg1$ copies~\cite{Demkowicz2020,Albarelli2020}.

Hitherto, the application of super-resolving measurements in imaging has focused primarily on ``beating'' the diffraction limit and 
maximising the precision in estimating the sources separation, $2d$, while assuming full control over all other parameters, in 
particular, the centroid's position, $x_{\mathrm{c}}$.  Of particular importance is the fact that the measurement that attains the 
quantum Fisher information when estimating \emph{only} the separation between two incoherent point sources is a projective 
measurement that does not depend on knowing $d$ in advance~\cite{Tsang2016}.  It does, however, require \emph{perfect} 
knowledge of the centroid, $x_{\mathrm{c}}$, of the PSF as well as perfect positioning of SPADE so that any  misalignment, 
$\delta\propto\theta$ in \eqnref{eq:parC}, can always be set to zero.

The separation can be estimated without requiring any knowledge about the centroid, if one has access to a quantum memory 
with a long coherence time so as to store photons collected during several independent experimental rounds ($n>1$) and be 
able to implement collective measurements~\cite{Ragy2016}. A proof-of-principle experiment that makes use of a 
measurement on a doublet of photons ($n=2$) and allows for simultaneous 
estimation of both the centroid and the separation of the sources has been reported recently~\cite{Parniak2018}. 
This has been achieved by encoding the spatial distribution of two incoherent sources into the spatial profile of 
a single photon generated in the laboratory. Utilising a pair of such photons and interfering them as in the
the Hong-Ou-Mandel experiment~\cite{HOM}, the information about both separation and centroid parameters can be
harmlessly retrieved, while estimating the former with precision beyond the diffraction limit~\cite{Parniak2018}. 
On the other hand, a recent theoretical study has proposed the use of direct imaging and SPADE techniques in 
parallel~\cite{Grace2019}. Direct imaging is performed repeatedly to part of the incoming radiation adjusting the exact position 
of SPADE via a servo feedback mechanism, in order to gradually reduce the misalignment, $\theta$ in Eq.~\eqref{eq:parC}, 
with increasing number of experimental repetitions. 

In the next subsection we use the qubit model to obtain the optimum measurement strategy for estimating the 
separation between two incoherent sources in presence of misalignment. 

\subsection{\label{sec:parestresults}Separation estimation under misalignment in the qubit approximation}

\begin{figure}[t!]  
   \centering
   \includegraphics[keepaspectratio, width=0.45\textwidth]{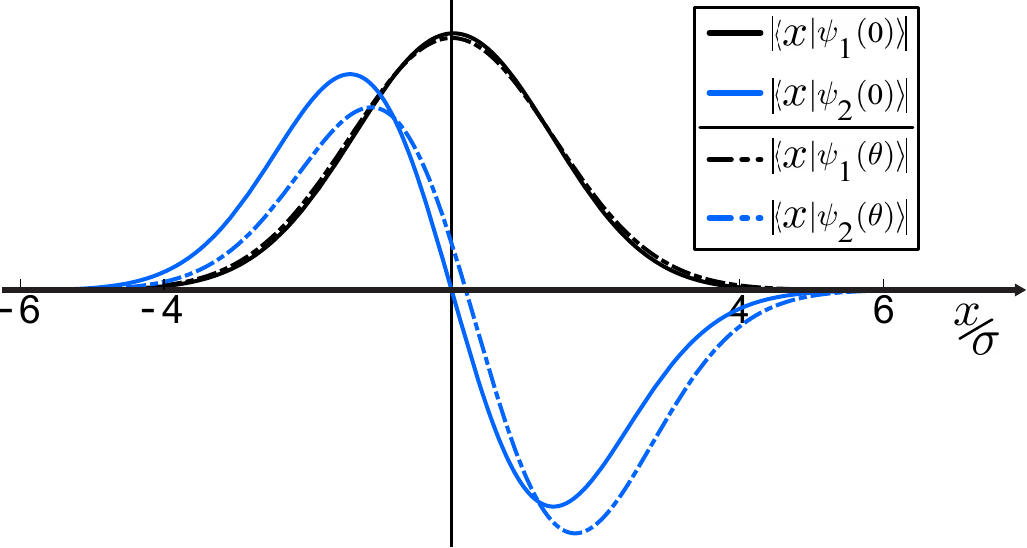}
   \caption{Spatial representation of the ROTADE operators for aligned (solid line, $\theta=0$) and misaligned (dot-
   dashed, $\theta=0.4$) measurement.  The plots are given for $\frac{d}{\sigma}=0.25$. $\ket{\psi_\alpha(\theta)}$ is defined in 
   Eq.~\eqref{eq:eigendecomp}. }
   \label{fig:meashelst}
\end{figure}
 
Assuming the separation between the incoherent sources to be small---as assured in the super-resolution regime---we use the 
qubit model in order to construct the optimal measurement for estimating the separation between two point sources under 
misalignment. We begin by first considering the Gaussian PSF.  The eigenvalues and corresponding eigenvectors 
of $\rho^{(2)}_\mathrm{G}$ are 
\begin{equation}
\begin{split}
\mu_1(\epsilon)=\frac{\epsilon^2}{4},\quad &
\ket{\psi_1(\theta)}=\sin\frac{\theta}{2}\ket{0}+\cos\frac{\theta}{2}\ket{1} \\
\mu_2(\epsilon)=1-\mu_1(\epsilon),\quad & \ket{\psi_2(\theta)}=-\cos\frac{\theta}{2}\ket{0}+\sin\frac{\theta}{2}\ket{1}.
\label{eq:eigendecomp}
\end{split}
\end{equation}
Using Eq.~\eqref{eq:sld} the corresponding SLD operators are, in the eigenbasis $\{\ket{\psi_1(\theta)},\ket{\psi_2(\theta)}\}$:
\begin{align}
\cL_\theta&= \left(1-\frac{\epsilon^2}{2}\right)\sx
,\qquad
\cL_\epsilon=\bematrix
\frac{2}{\epsilon} &0\\
0& \frac{2\epsilon}{\epsilon^2-4}
\ematrix.
\label{eq:QMSLDs}
\end{align}
Observe that $[\cL_\theta,\cL_\epsilon]\neq 0$, meaning that the optimal measurements for each of these parameters are 
incompatible.  However,  $\tr\left(\rho^{(2)}_\mathrm{G}[\cL_\theta,\cL_\epsilon]\right)=0$, which implies that there exists a 
possibly joint measurement on all $n$ photons that saturates the quantum Cram\'er-Rao bound given by 
\begin{equation}
\langle(\hat{\theta}-\theta,\hat{\epsilon}-\epsilon)^T(\hat{\theta}-\theta,\hat{\epsilon}-\epsilon)\rangle\geq \frac{1}{n}
\bematrix
\frac{1}{1-\epsilon^2} & 0\\
0& \frac{1}{1+\frac{\epsilon^2}{4}}
\ematrix.
\label{eq:QMCRbound}
\end{equation}
The eigenvectors of the SLD operators (Eq.~\eqref{eq:QMSLDs}) are given by 
\begin{align}
\ket{\theta_\pm}&=\frac{1}{\sqrt{2}}\left(\left(\sin\frac{\theta}{2}\pm\cos\frac{\theta}{2}\right)\ket{0}+\left(\sin\frac{\theta}
{2}\mp\cos\frac{\theta}{2}\right)\ket{1}\right),
\label{eq:sldiegenvecs_misal}
\\
\ket{\epsilon_{\alpha}}&=\ket{\psi_\alpha(\theta)}\quad\text{with}\quad\alpha\in\{1,2\},
\label{eq:sldiegenvecs_sepa}
\end{align}
respectively. As $\cL_\epsilon$ is a diagonal operator, the optimal measurement in Eq.~\eqref{eq:sldiegenvecs_sepa} for 
estimating the re-scaled separation $\epsilon$ between the two sources according to the qubit model is simply given by a 
projective measurement in the eigenbasis of Eq.~\eqref{eq:eigendecomp}. Henceforth, we shall refer to this measurement as the 
\emph{rotated mode demultiplexer} (ROTADE), i.e.~the detection scheme depicted schematically in Fig.~\ref{fig:ROTADE} with 
the rotation $R(\delta)$ adequately adjusted to $\delta=\theta\sigma$.

\begin{figure}[t!]  
   \centering
   \includegraphics[width=0.45\textwidth]{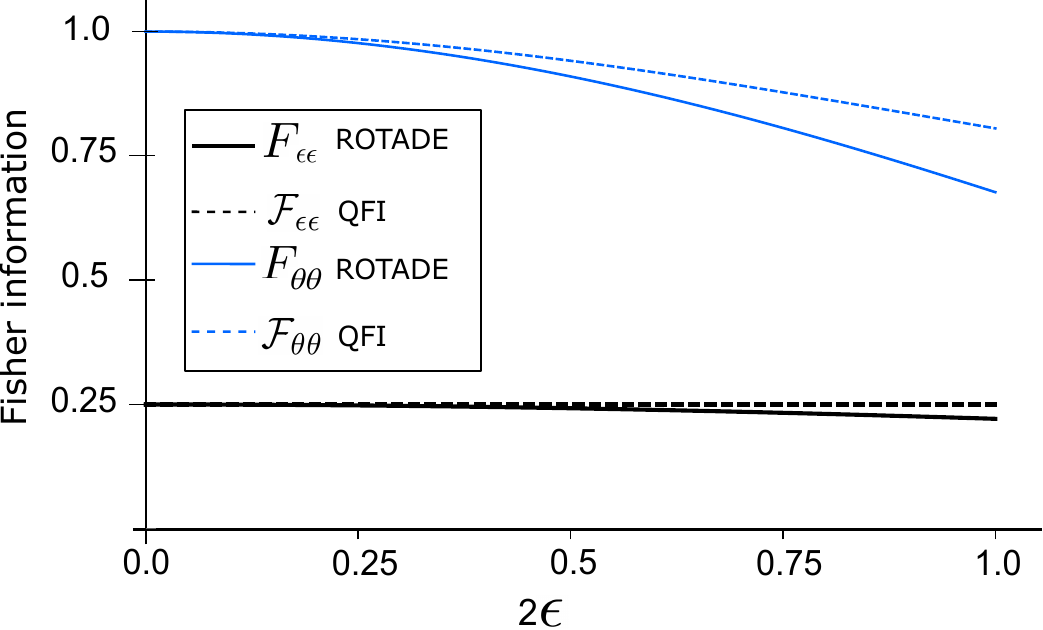}
   \caption{\emph{Gaussian aperture:} Quantum Fisher information, $\frac{\cF_{\bm{\lambda}}}{n}$, attained in \cite{Tsang2016} 
   (dashed lines) and classical Fisher information associated to the ROTADE measurement $\frac{F_{\bm{\lambda}}}{n}$ (solid 
   lines), for separation $\epsilon$ (red) and misalignment $\theta$ (blue) parameters for perfect alignment $\theta=0$, as a 
   function of $2\epsilon$. As the POVM Eq.~\ref{eq:sldiegenvecs_sepa} is derived based on the qubit model, it ceases to be 
   optimal with increasing separation of the sources (here for $\epsilon\lesssim 0.1$).}
   \label{fig:TsangQepz}
\end{figure}

In order to compare the quality of the ROTADE measurement, we use Eq.~\eqref{eq:Taylor_approx} to map the 
measurement operators into their position-based representation. The latter are shown in Fig.~\ref{fig:meashelst}. One can 
then explicitly determine the probability distribution arising from these measurements and hence the corresponding Fisher 
information using Eq.~\eqref{eq:Classical_Fisher}. The results are shown in Fig.~\ref{fig:TsangQepz}, where we compare the 
performance of ROTADE with the quantum Fisher information~\cite{Tsang2016} for $\theta=0$, i.e.~in the absence of 
misalignment. We see that up to separations $\epsilon=\frac{d}{\sigma}\lesssim0.5$ the Fisher information of
ROTADE drops to $\approx90\%$ of the optimal value.  On the other hand, up to 
$\epsilon=\frac{d}{\sigma}\lesssim0.1$ ROTADE maintains its optimality, emphasising that the qubit model 
approximates well the super-resolution problem in this regime. Hence, in the limit where the qubit model holds, counting photons 
only in the first two HG modes suffices to estimate the separation.

\begin{figure}[t!]  
   \centering
   \includegraphics[width=0.43\textwidth]{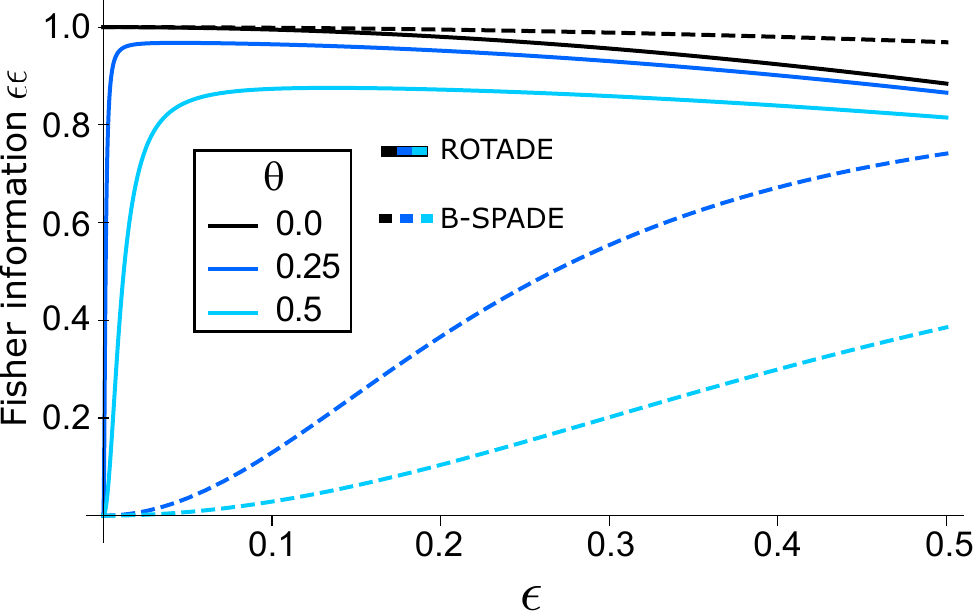}
   \caption{\emph{Gaussian aperture:} B-SPADE Fisher information attained in~\cite{Tsang2016} (dashed lines) and ROTADE 
   Fisher information (solid lines) for separation estimation under misalignment, $\frac{F_{\epsilon\epsilon}}{n}$, as a function of 
   the separation.}
   \label{fig:Fisherqubit}
\end{figure}

A simpler measurement that also achieves the quantum bound (Fig.~\ref{fig:TsangQepz}) is B-SPADE~\cite{Tsang2016}. This is a 
coarse grained version of SPADE where only photons in the fundamental HG mode of SPADE are counted, while lumping all 
other modes to produce a single photon-count outcome. As the probability of detecting the zeroth HG mode occurs regardless of the separation, $d$, B-SPADE is more experimentally friendly, but suffers in the same manner as SPADE from the misalignment 
problem. In Fig.~\ref{fig:Fisherqubit} we compare the performance of ROTADE with B-SPADE in estimating the separation under 
a misalignment $\theta\leq 0.5$. 

In order to capture the difference between the aforementioned measurements we compare the Taylor expansions of their 
corresponding Fisher information up to first non-trivial order, for small separation $\epsilon$.  These are given by 
\begin{equation}
\begin{split}
F^{(\mathrm{R})}_{\bm{\epsilon\epsilon}}(\epsilon)&\approx \epsilon^2 C^{(R)}(\theta)\\
F^{(\mathrm{B})}_{\bm{\epsilon\epsilon}}(\epsilon)&\approx \epsilon^2 C^{(B)}(\theta),
\end{split}
\label{eq:TSeriesFishers}
\end{equation}
where $C^{(R)}(\theta),\, C^{(B)}(\theta)$ are coefficients pertaining to the measurements themselves and depend only on the 
misalignment $\theta$ (notably they are independent of $n$ and $\epsilon$).  The behaviour of these coefficients governs the 
precise minimal resolvable distance for each measurement as we now explain.

The signal-to-noise-ratio $\epsilon/\Delta\epsilon$ can be expressed as 
\begin{equation}
\epsilon\sqrt{n F^{(\mathrm{\#})}_{\bm{\epsilon\epsilon}}(\epsilon)}\geq 1,
\label{eq:sig2noise}
\end{equation}
where $\#\in(R,B)$.  The minimal resolvable separation, $\epsilon^{(\#)}_{\min}(\theta)$, for each measurement is defined as 
that $\epsilon$ in Eq.~\eqref{eq:sig2noise} for which equality holds.  Using the approximations of Eq.~\eqref{eq:TSeriesFishers} 
one obtains 
\begin{equation}
\epsilon^{(\#)}_{\min}(\theta)=\frac{1}{\sqrt[4]{n C^{(\#)}(\theta)}}. 
\label{eq:epsilonminsnr}
\end{equation}
Taylor expanding the functions $C^{(\#)}(\theta)^{-1}$ to first non-trivial order in $\theta$ one obtains 
\begin{equation}
\begin{split}
C^{(\mathrm{R})}(\theta)^{-1}&\approx\frac{\theta^6}{12^2}\\
C^{(\mathrm{B})}(\theta)^{-1}&\approx \theta^2.
\end{split}
\label{eq:coefTS}
\end{equation}
It follows that 
\begin{equation}
\begin{split}
\epsilon^{(\mathrm{R})}_{\min}(\theta)&\approx \frac{1}{\sqrt[4]{n}}\frac{\theta^{\frac{3}{2}}}{\sqrt{12}}\\
\epsilon^{(\mathrm{B})}_{\min}(\theta)&\approx \frac{\sqrt{\theta}}{\sqrt[4]{n}},
\end{split}
\label{eq:epsilonmins}
\end{equation}
where $\epsilon_{\min}(\theta)\propto n^{-\frac{1}{4}}$ is a consequence of $F_{\bm{\epsilon\epsilon}}\propto\epsilon^2$ 
in Eq.~\eqref{eq:TSeriesFishers}. In contrast,
observe that in the ideal case of no misalignment, for which $F_{\bm{\epsilon\epsilon}}\propto 1$, the minimal resolvable distance 
scales as $\epsilon_{\min}(0)\propto n^{-\frac{1}{2}}$.

The quadratic increase in the scaling  of $\epsilon_{\min}$ for both ROTADE and B-SPADE due to misalignment mimicks closely
the behaviour of cross-talk between the measurement modes addressed recently by Guessner {\it et al.}~\cite{gessner2020}.  
As our qubit approximation puts us in the regime of only monitoring the first two HG modes, and misalignment corresponds to a 
unitary rotation of the latter, it follows that this unitary rotation can be interpreted as the cross-talk matrix of~\cite{gessner2020}.
As the cross-talk probability between the two modes is proportional to $\sin^2\theta\approx\theta^2$, $\epsilon^{(B)}_{\min}$ of 
Eq.~\eqref{eq:epsilonmins} follows precisely the analytical model for uniform cross-talk of~\cite{gessner2020}. 

Our results show that super-resolution is impossible if the initial demultiplexing of the incoming radiation suffers any misalignment, 
even if the latter is known.  Nevertheless, cross-modulation techniques between the two primary HG modes can help in 
significantly reducing the minimum resolvable distance.

In Appendix~\ref{appendix:Sinc} we obtain the optimal measurement under misalignment for the Sinc PSF, as well as the 
minimum resolvable distance.  Our results confirm the efficacy of the qubit model; for whatever PSF the first two modes are the 
most relevant ones in estimating the position of light sources with separation well below the diffraction limit. In the next section, 
we will discuss how the optimal measurement under misalignment derived using the qubit model is also optimal for the task of 
discriminating whether the incoming radiation is due to two incoherent point sources or one source with twice the power under 
misalignment.

\section{\label{sec:hypothesistesting} Classical and quantum state discrimination: one or two point sources.}

Hitherto our focus was to estimate the relevant parameters of two incoherent point sources.  However, a more pertinent question 
is whether the incoming radiation is due to two incoherent point sources very close together (\emph{the two source hypothesis, 
$H^{(2)}$}), or one point source with twice the power (\emph{the one source hypothesis, $H^{(1)}$}). To that end we first review 
the fundamentals of classical and quantum decision theory and, in particular, simple binary hypothesis 
testing~\cite{Helstrom1976,Holevo1982}. We then apply these tools to optimally discriminate between $H^{(1)}, \, H^{(2)}$ in the 
presence of misalignment and compare the performance of ROTADE with measurements in the literature, showing that our 
measurement outperforms all the latter.

\subsection{\label{sec:Hintro} Classical and quantum hypothesis testing}

A fundamental problem in decision theory is to discriminate among several possible hypothesis based on a number, $n$, of 
observations. The simplest such scenario---known as binary hypothesis testing---occurs when there are two hypothesis, 
$H^{(1)},\, H^{(2)}$ that need to be discriminated.  For simplicity, assume that each observation consists of a finite set of possible 
outcomes $y\in Y$~\footnote{The case of continuous random variables follows similarly}.  Under hypothesis $H^{(i)}$, these 
outcomes are distributed according to $p(y|H^{(i)})$, and thus the problem becomes one of determining from which probability 
distribution the random variable $Y$ is drawn.

For a single observation ($n=1$) let $f:Y\to\{H^{(1)},\,H^{(2)}\}$ be a decision rule.  Under such a decision rule the probability of 
making an error based on a single observation is  
\begin{equation}
P_{\mathrm{err}}=\frac{1}{2}\Big(p\left(f(y)=H^{(2)}|H^{(1)}\right)+p\left(f(y)=H^{(1)}|H^{(2)}\right)\Big),
\label{eq:classicalHT}
\end{equation}
where we have assumed that each hypothesis is equally likely. The conditional probabilities 
$p\left(f(y)=H^{(2)}|H^{(1)}\right),\, p\left(f(y)=H^{(1)}|H^{(2)}\right)$ are the \emph{type-1} (mistaking one source for two) and 
\emph{type-2} (mistaking two sources for one) errors, respectively. For binary hypothesis testing, the optimal decision rule is to 
assign the hypothesis with the highest posterior distribution~\cite{Renyi1966,Jerzy1933} which, for equally likely hypothesis, 
translates to 
\begin{equation}
f(y)=\begin{cases}
H^{(1)} & \mathrm{if}\quad p\left(y|H^{(1)})>p(y|H^{(2)}\right)\\
H^{(2)} & \mathrm{if}\quad p\left(y|H^{(2)})>p(y|H^{(1)}\right)\\
\mathrm{any}&\mathrm{if}\quad p\left(y|H^{(1)})=p(y|H^{(2)}\right),
\end{cases}
\end{equation}
and the corresponding probability of error reads 
\begin{align}\nonumber
P_{\mathrm{err}}&=\sum_{y\in Y} p(y)\,\mathrm{min} \left\{p\left(H^{(i)}|y\right)\right\} \\    \nonumber
&=\sum_{y\in Y}\mathrm{min} \left\{p\left(y,H^{(i)}\right)\right\} \\ 
&=\frac{1}{2}\left(1-\frac{1}{2}\sum_{y\in Y}\bigg\lvert \,p\left(y|H^{(1)}\right)-p\left(y|H^{(2)}\right)\bigg\rvert\right)
\label{eq:prob_err}
\end{align}
where we have made use of the identity $\mathrm{min}\{a,b\}=\frac{1}{2}\left(a+b-\lvert a-b\rvert\right)$ in order to obtain the last 
equality.

Quantum hypothesis testing now follows by noting that $p\left(y|H^{(i)}\right)=\tr\left(E_y\,\rho^{(i)}\right)$ where $\{E_y\}$ 
constitute a POVM and the hypothesis, $\rho^{(i)},\, i\in(1,2)$, are given by Eqs.~(\ref{eq:single_source},~\ref{eq:two_sources}). 
Doing the appropriate substitutions in Eq.~\eqref{eq:prob_err} one obtains
\begin{equation}
P_{\mathrm{err}}=\frac{1}{2}\left(1-\frac{1}{2}\tr\left(\sum_{y\in Y}E_y \bigg\lvert \rho^{(1)}-\rho^{(2)}\bigg\rvert\right)\right).
\label{eq:quantum_error}
\end{equation}

Unlike the classical case, in quantum binary hypothesis testing we are free to choose among all admissible POVMs the one that 
yields the smallest probability of error.  The optimal measurement in this case was derived by Helstrom~\cite{Helstrom1976} and 
corresponds to a two outcome measurement $\{E_0,\, E_1\}$ on the positive and negative eigenspaces of the operator 
\begin{equation}
\Gamma:=\frac{1}{2}\left(\rho^{(2)}-\rho^{(1)}\right).
\label{eq:Helstrom}
\end{equation} 
Given $n$ copies of the initial state, the Helstrom measurement is generally a collective measurement on the positive and 
negative eigenspaces of $\Gamma^{\otimes n}=\frac{1}{2}\left(\rho^{(2)\otimes n}-\rho^{(1)\otimes n}\right)$.  For clarity we shall 
call the single copy optimal measurement as the \emph{Helstrom measurement}, and the overall optimal measurement on $n$ 
copies as the \emph{collective Helstrom measurement}.

The probability of error decreases exponentially with the number of copies $n$.  In order to compare the performance of different 
measurement strategies one needs to determine the rate at which this error probability decreases. For an asymptotically large 
($n\rightarrow\infty$) number of observations the probability of error saturates Chernoff's inequality~\cite{Chernoff1952}:
\begin{equation}
P_{\mathrm{err}}(n) \leq e^{-n\xi},
\label{eq:Chernoff_bound}
\end{equation}
where,
\begin{equation}
\xi:=-\log\min_{0\leq s \leq 1}\sum_{y\in Y} p\left(y|H^{(1)}\right)^{s}p\left(y|H^{(2)}\right)^{1-s}
\label{eq:cher}
\end{equation}
is the Chernoff exponent. In the case of quantum hypothesis testing, the asymptotic error rate is given by the quantum Chernoff 
exponent~\cite{Audenaert2007}:
\begin{equation}
\xi\le\xi^{(QM)}:=-\log\min_{0\leq s \leq 1}\tr\!\left\{\left(\rho^{(1)}\right)^{s}\left(\rho^{(2)}\right)^{1-s}\right\},
\label{QChernoff}
\end{equation}
which is generally larger than its classical counterpart.  Note that the quantum Chernoff exponent only depends on the quantum 
states to be discriminated, and is independent of the measurement performed.  Nonetheless, the inequality in 
Eq.~\eqref{eq:Chernoff_bound} is asymptotically achievable in the limit of infinite copies. In this limit, $\xi$ reaches the 
ultimate quantum bound $\xi^{(QM)}$ of asymptotic (symmetric) hypothesis testing~\cite{Nussbaum2009}. 
However, such attainability may require a collective Helstrom measurement to be performed on all the $n\to\infty$ copies.

Surprisingly, it was already Helstrom~\cite{Helstrom1973} who first addressed the problem of discriminating one-vs-two 
incoherent point sources of light with tools from hypothesis testing, and derived a sub-optimal measurement that; (i) lacks a 
physical realisation and (ii) requires knowledge of the separation of the two sources.  Krovi {\it et al.}~\cite{Krovi2016} derived 
the optimal quantum mechanical measurement that achieves the quantum Chernoff bound for the case where the separation of 
the two point sources is known and showed how to experimentally implement it.  Shortly after, \cite{Lu2018} showed that the 
B-SPADE measurement of~\cite{Tsang2016} achieves the quantum Chernoff bound for one-vs-two sources of \emph{arbitrary} 
separation.  However, just like in the estimation case, all these works assumed that the centre of the single source, as well as the 
centroid of the two source hypothesis, to be perfectly aligned with the demultiplexing measurements and neglected any noise at 
the detectors.

In the next subsection we analyse the behaviour of B-SPADE under misalignment and show that it falls short of the quantum 
optimal Chernoff bound. Using the qubit model we derive an alternative measurement strategy that is also sub-optimal but 
outperforms the B-SPADE under misalignment by far.

\subsection{\label{sec:Hqubit} State discrimination in the qubit approximation---the Helstrom measurement}

Our aim is to determine whether the PSF observed at the misaligned imaging system is due to two incoherent point sources of 
equal intensities or a single source with twice the intensity.  For the remainder of this section, we shall work with the Gaussian 
PSF (results for the Sinc PSF can be derived in a similar fashion and are presented in Appendix~\ref{appendix:ChenoffSinc}). 
Using the qubit model the matrix $\Gamma$ of Eq.~\eqref{eq:Helstrom} can be explicitly computed to be
\begin{equation}
\Gamma= \frac{1}{4}
   \begin{pmatrix}
-\cos\theta_{\mathrm{0}}-\frac{1}{2}\cos\theta_{\mathrm{c}}(\epsilon^{2}-2) & \sin\theta_{\mathrm{0}}+\frac{1}{2}\sin\theta_{\mathrm{c}}(\epsilon^{2}-2) \\
    \\
    \sin\theta_{\mathrm{0}}+\frac{1}{2}\sin\theta_{\mathrm{c}}(\epsilon^{2}-2) & \cos\theta_{\mathrm{0}}+\frac{1}{2}\cos\theta_{\mathrm{c}}(\epsilon^{2}-2)
  \end{pmatrix},
  \label{eq:gamma}
\end{equation}
where $\theta_{\mathrm{0}}=\frac{x_{\mathrm{0}}-x_R}{\sigma}$ is the misalignment relative to the centre of a single source PSF, 
$\theta_{\mathrm{c}}=\frac{x_{\mathrm{c}}-x_R}{\sigma}$ is the misalignment relative to the centroid, $x_{\mathrm{c}}$, of the two 
sources PSF, and $\epsilon$ is defined as in Eq.~\ref{eq:parC}.  Notice that, in principle, the centre of a single source need not 
coincide with the centroid of two sources, nor with the position of the demultiplexing measurement, 
$x_{\mathrm{0}}\neq x_{\mathrm{c}}\neq x_R \quad(\theta_{\mathrm{0}}\neq\theta_{\mathrm{c}})$. Nonetheless, hereafter we 
shall restrict our analysis to the case where only the demultiplexing measurements are misaligned, hence we will define: 
\begin{equation}
\theta:=\theta_{\mathrm{0}}=\theta_{\mathrm{c}}.
\end{equation}
In this regime, the Helstrom measurement is independent of separation and is equivalent to ROTADE.
 
\begin{figure}[ht!]  
   \centering
   \includegraphics[keepaspectratio, width=0.45\textwidth]{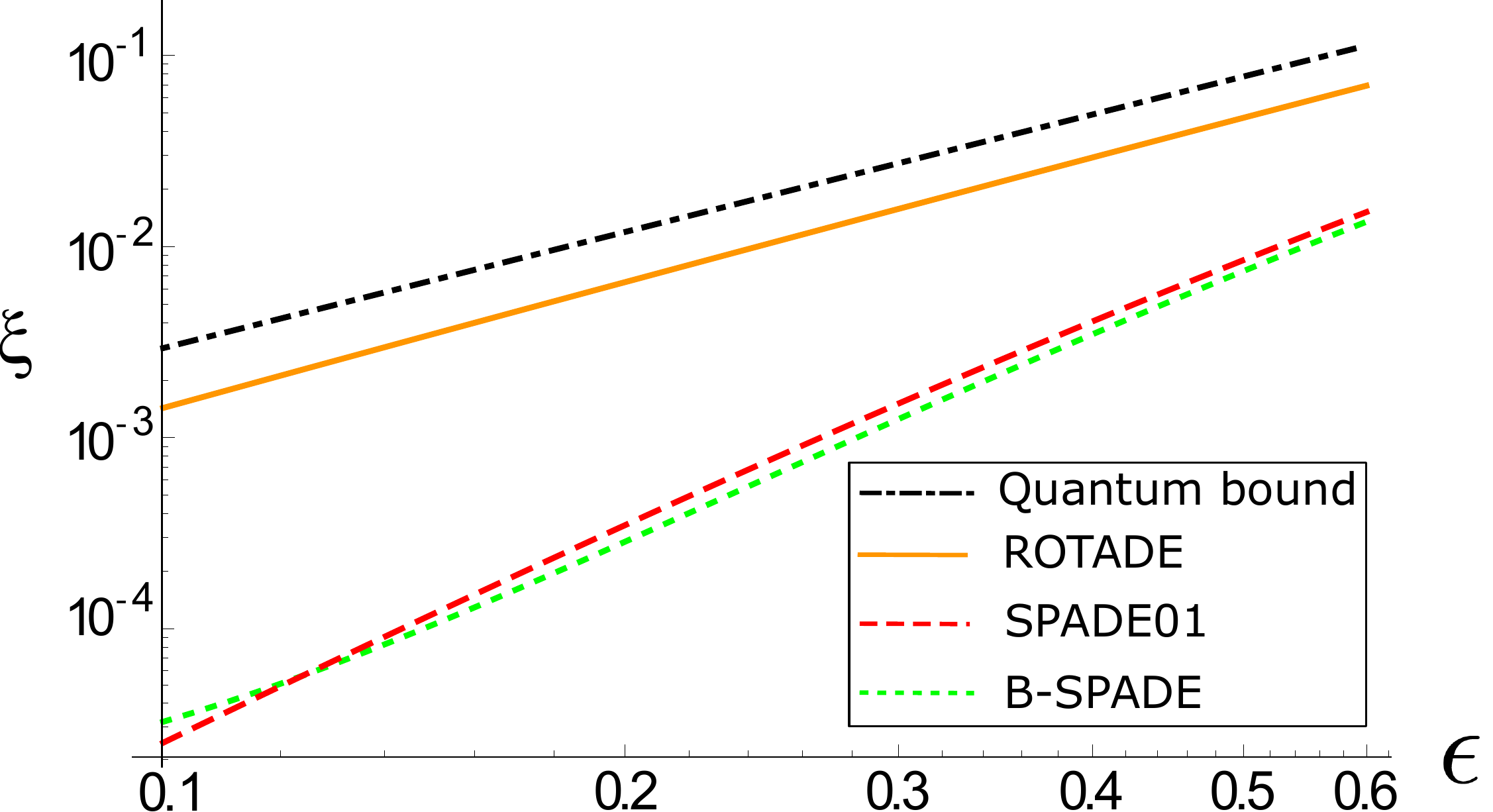}
   \caption{Numerical optimisation of the Chernoff exponent under misalignment as a function of the separation in $\log-\log$ plot, 
   for $x_R=0.4$ and $\sigma=1$.}
   \label{fig:HGvsH}
\end{figure}  
 
In case the detector and centroid are perfectly aligned, $\theta=0$, ROTADE is only the projection onto the zeroth and first HG 
modes.  We shall refer to this measurement as SPADE$01$, in order to distinguish it from B-SPADE which projects 
only on the zeroth mode. We remark that all measurement strategies reach the quantum bound for zero misalignment. The main 
advantages of SPADE$01$ for aligned measurement device are:  it is independent of the two-sources separation, the need to 
count photons only in the first two HG modes (photons coupling to higher modes correspond to no-clicks and are insignificant to 
the measurement statistics), and the unambiguous two-source discrimination whenever a photon is detected in the first HG mode.  
These results are shown in Appendix~\ref{appendix:Qubitperformance}. 

\begin{table}[ht!]  
	\begin{center}

		\resizebox{0.48\textwidth}{!}{
		\begin{tabular}{l||c|c|}  
			Measurement & $p(f(y)=H^{(2)}|H^{(1)})$  & $p(f(y)=H^{(1)}|H^{(2)})$ \\
			\hline
			\hline
			 ROTADE & $\frac{\theta^6}{576 \sigma^6}$     &  $\exp{\left(-\frac{d^2}{4\sigma^2}\right)}\left(1-\frac{d^2 \theta^2}
			 {16\sigma^4}\right)$\\
			 \hline
			 SPADE01 & $\frac{\theta^2}{4\sigma^2}$ &  $\exp{\left(-\frac{d^2}{4\sigma^2}\right)}\left(1+\frac{(d^2-2\sigma^2) 
			 \theta^2}{8\sigma^4}\right)$\\
			\hline
			B-SPADE & $\frac{d^2\theta^2\csch\left(\frac{d^2}{4\sigma^2}\right)}{16\sigma^4}$ &  $\exp{\left(-\frac{d^2}
			{4\sigma^2}\right)}\left(1+\frac{(d^2-2\sigma^2) \theta^2}{8\sigma^4}\right)$\\
			\hline
			\end{tabular}
			}
		\end{center}
		\caption{Taylor expansion to the first non-trivial order in $\theta$ for the $type-1$ (second column) and $type-2$ (third 
		column) error probabilities for ROTADE, SPADE$01$ and B-SPADE.}
		\label{table:2}	
\end{table}

Table~\ref{table:2} shows how the one shot error probability scales as a function of the misalignment for the first 
non-trivial order of the Taylor expansion around $\theta=0$. Notice that for ROTADE, the 
$type-1$ error, responsible for the unambiguous determination of the two-source hypothesis, is four orders of 
magnitude smaller compared to that of SPADE$01$ and  B-SPADE.  Hence in the single-shot scenario ROTADE 
significantly outperforms both these measurements.

\begin{figure}[ht!]  
   \centering
   \includegraphics[keepaspectratio, width=0.45\textwidth]{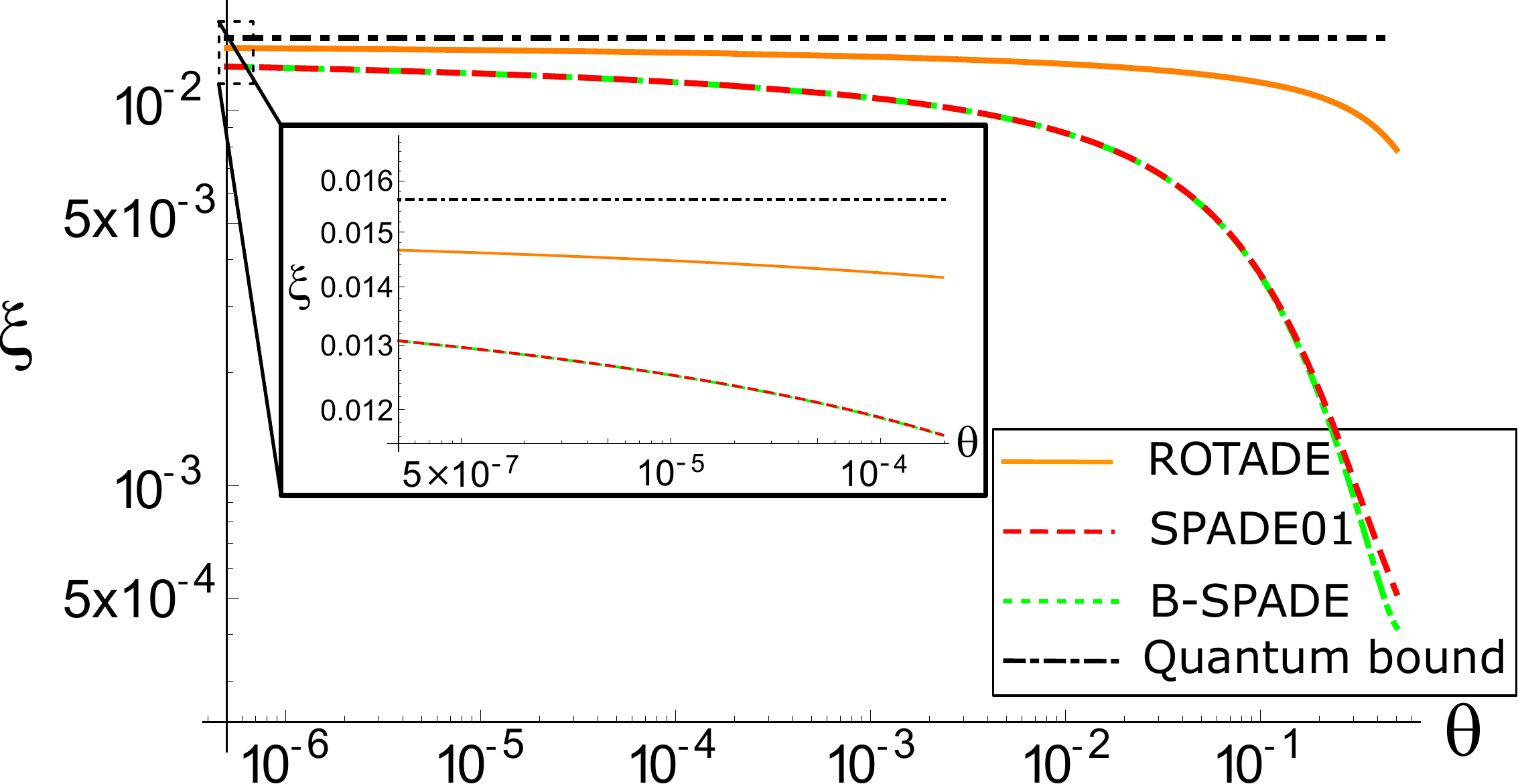}
   \caption{Numerical determination of the Chernoff exponent as a function of the misalignment, for separation $\epsilon=0.25$ 
   between a pair of sources with $\sigma=1$. The inset shows how the Chernoff exponent varies for the three relevant 
   measurement strategies for $\theta\approx0$.}
   \label{fig:HGvsHxr}
\end{figure}

The Chernoff exponent of the SPADE$01$ measurement under misalignment behaves similarly to that of B-SPADE, the 
asymptotic results of all measurement strategies under misalignment as function of separation are represented in 
Fig.~\ref{fig:HGvsH}. However, in contrast with the aligned scenario, for $\theta\neq 0$ the probability of detecting photons into 
higher HG modes is non-negligible, and corresponds to the no-click probability. This probability represents the intrinsic error of the 
qubit model and it increases with misalignment (for details see Appendix~\ref{appendix:Qubitperformance}). 

Unfortunately, we are unable to obtain an analytic expression for the Chernoff exponent under misalignment for any of the three 
strategies.  This is because the $s$ that minimises the Chernoff exponent in Eq.~\ref{eq:cher} explicitly depends on $\theta$. 
Fig.~\ref{fig:HGvsHxr} presents a numerical optimisation for the Chernoff exponent as a function of the misalignment. We observe 
that for all $\theta>0$ ROTADE outperforms both SPADE$01$ and B-SPADE, which is to be expected as ROTADE includes the 
knowledge on the amount of misalignment. Nonetheless, for \emph{exactly} $\theta=0$ all the corresponding Chernoff exponents 
coincide with the quantum bound, what manifests their discontinuity as $\theta\to0_+$.

\section{\label{sec:conclusion} Conclusions}

In this work we have analysed the impact of a {\it de facto} misalignment in the demultiplexing measurements of an optical 
imaging system in both estimating the separation of incoherent light sources, as well as discriminating between one-versus-two 
incoherent point sources. By allowing for linear-optical post processing of the two dominant demultiplexed modes, we have shown 
that super-resolution cannot be perfectly restored, even if the value of the misalignment is {\it a priori known}.  Using quantum
information methods, we have constructed misalignment-dependent  strategies that, in the case of estimation, allow for subdiffraction-limited estimation of the separation of two incoherent light sources and analytically determined the dependence of both 
the estimation precision as well as the minimal resolvable distance as a function of the misalignment.  Remarkably, the same 
measurement exhibits improved performance also in the task of discriminating among the one-vs-two source hypothesis showing 
significant improvement in both the single-shot as well as asymptotic probability of error.




Several interesting questions still remain.  How does misalignment affect estimation precision when both the separation as well as 
the relative intensities of the two incoherent point sources need to be estimated?  In the case of discrimination an interesting 
question occurs when the centres of the two hypothesis do not coincide, i.e., $x_0-x_{\mathrm{c}}\ll \sigma^2$ but neither $x_0$ 
nor $x_{\mathrm{c}}$ coincides with $x_R$.  Then, the optimal Helstrom measurement \emph{does} depend on knowing the 
separation between the two sources and it remains an open question if there exists a classical measurement with super-resolving 
power.  We hope to answer these questions in the future. 

\section*{Acknowledgments}
We thank Wojciech Wasilewski and Micha\l{} Parniak for helpful comments. JOA and ML acknowledges support from ERC AdG NOQIA, Spanish Ministry of Economy and Competitiveness (``Severo Ochoa'' program for Centres of Excellence in R$\&$D (CEX2019-000910-S), Plan National FIDEUA PID2019-106901GB-I00/10.13039 / 501100011033, FPI), Fundaci\'{o} Privada Cellex, Fundaci\'{o} Mir-Puig, and from Generalitat de Catalunya (AGAUR Grant No. 2017 SGR 1341, CERCA program, QuantumCAT U16-011424, co-funded by ERDF Operational Program of Catalonia 2014-2020), MINECO-EU QUANTERA MAQS (funded by State Research Agency (AEI) PCI2019-111828-2 / 10.13039/501100011033), EU Horizon 2020 FET-OPEN OPTOLogic (Grant No 899794), and the National Science Centre, Poland-Symfonia Grant No. 2016/20/W/ST4/00314. JK is supported by the Foundation for Polish Science under the ``Quantum Optical Technologies'' project carried out within the International Research Agendas programme, co-financed by the European Union under the European Regional Development Fund. CH acknowledges financial support from the VILLUM FONDEN via the QMATH Centre of Excellence (Grant no. 10059). MS acknowledges support from Spanish MINECO reference FIS2016-80681-P (with the support of AEI/FEDER,EU); the Generalitat de Catalunya, project CIRIT 2017-SGR-1127 and the Baidu-UAB collaborative project `Learning of Quantum Hidden Markov Models'.

\appendix
\section*{Appendix}
\section{Estimating the separation between Sinc-Bessel modes under misalignment}
\label{appendix:Sinc}
In this appendix section, we present the results of estimating the separation between two incoherent point sources imaged by a system 
with a rectangular aperture. The PSF of such a system is given by the Sinc function (see Eq.~\eqref{eq:single_source_PSFb}).

\begin{figure}[ht!]  
   \centering
   \includegraphics[keepaspectratio, width=8.5cm]{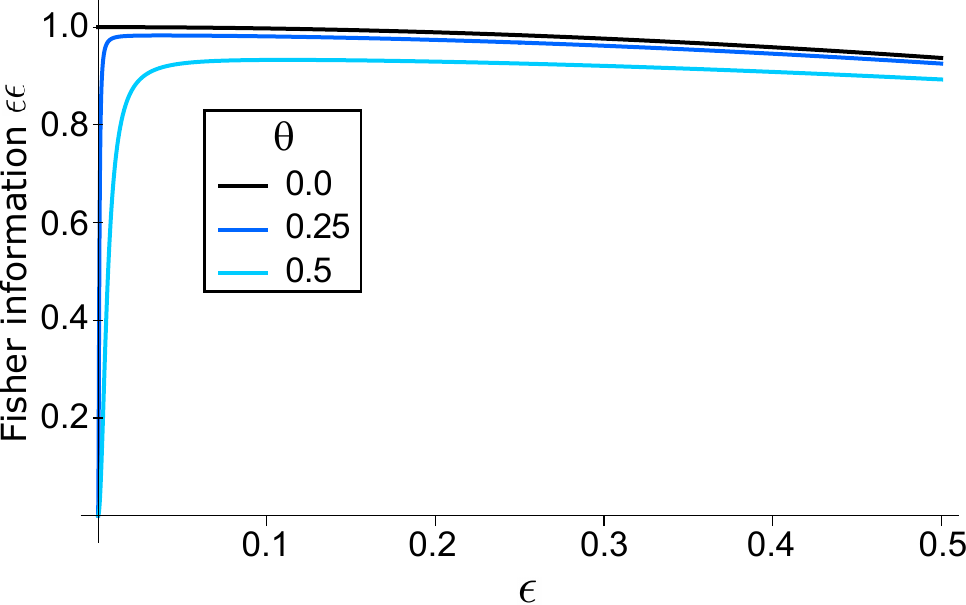}
   \caption{ \emph{Rectangular aperture:}
   $\frac{\cF_{\epsilon\epsilon}}{n}$ for the equivalent to ROTADE measurement, as a function of separation for a rectangular aperture with $\sigma=1$ for different misalignments.}
   \label{fig:QubitMultiControidSinc}
\end{figure}

Repeating the calculation in Section~\ref{sec:CramerRao} the eigenvalues and corresponding eigenvectors of $\rho^{(2)}_{\mathrm{S}}$ are:
\begin{equation}
\begin{split}
\mu_1(\epsilon)=\frac{\epsilon^2}{3},\quad &
\ket{\psi_1(\theta)}=\sin\frac{\theta}{\sqrt{3}}\ket{0}+\cos\frac{\theta}{\sqrt{3}}\ket{1}\\
\mu_2(\epsilon)=1-\mu_1(\epsilon),\quad & \ket{\psi_2(\theta)}=-\cos\frac{\theta}{\sqrt{3}}\ket{0}+\sin\frac{\theta}{\sqrt{3}}\ket{1},
\label{eq:eigendecompsinc}
\end{split}
\end{equation}
and using Eq.~\eqref{eq:sld} the corresponding SLD operators are, in the eigenbasis 
$\{\ket{\psi_1(\theta)},\ket{\psi_2(\theta)}\}$ are given by 
\begin{align}\nonumber
\cL_\theta&= \left(\frac{6-4\epsilon^2}{3\sqrt{3}}\right)\sx\\
\cL_\epsilon&=\frac{2}{\epsilon}\bematrix
1 &0\\
0& \frac{\epsilon^2}{\epsilon^2-3}
\ematrix.
\label{eq:QMSLDsinc}
\end{align}
The eigenvectors of the SLD operators can now easily be computed to be:
\begin{align}\nonumber
\ket{\theta_\pm}=&\frac{1}{\sqrt{2}}\left(\left(\sec\frac{2 \theta}{\sqrt{3}} \pm \tan\frac{2 \theta}{\sqrt{3}}\right) \sqrt{ 1 \mp \sin\frac{2 \theta}{\sqrt{3}}})\ket{0}+\right.\\
& \left.+\sqrt{ 1 \mp \sin\frac{2 \theta}{\sqrt{3}}})\ket{1}\right)\\
\ket{\epsilon_{\alpha}}=&\ket{\psi_\alpha(\theta)},
\label{eq:sldsinceigenvecs}
\end{align}
The optimal measurement to detect the separation for known misalignment is analogous to ROTADE with angle $\frac{\theta}{\sqrt{3}}$.

In the case of no misalignment the optimal measurement is given by the first two Sinc-Bessel modes~\cite{Kerviche2017}. In the presence of misalignment the optimal measurements furnished by the qubit model are unitarily related to the same two Sinc-Bessel modes. The Fisher information for various values of misalignment for the Sinc PSF are shown in Figure~\ref{fig:QubitMultiControidSinc}. Similar to Sec.~\ref{sec:parestresults} we can analyse the minimal resolvable distance under misalignment to estimate the separation of Sinc PSF $\epsilon^{(R_{\mathrm{Sinc}})}_{\min}(\theta)\approx\frac{\theta^{3/2}}{\sqrt[4]{n} \sqrt{5\sqrt{3^3}}}$, this is an improvement in contrast with the minimal resolvable distance of SPADE01 $\epsilon^{(01_{\mathrm{Sinc}})}_{\min}(\theta)\approx\frac{\sqrt[4]{3} \sqrt{\theta}}{\sqrt{2}}$.

\section{Discrimination of Sinc-Bessel modes}
\label{appendix:ChenoffSinc}

\begin{figure}[h!]  
   \centering
   \includegraphics[scale=0.38]{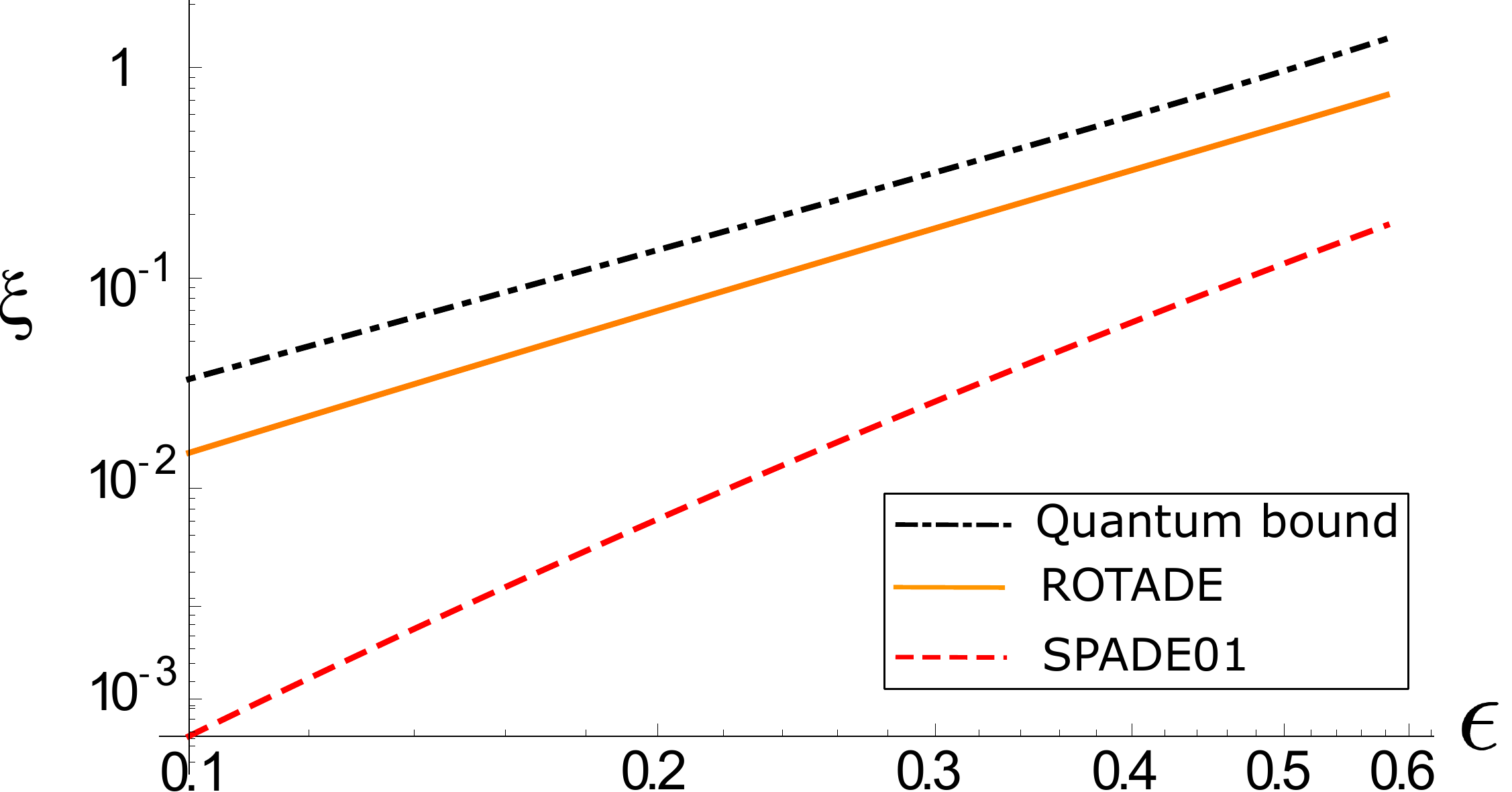}
   \caption{Chernoff exponent for Sinc PSF in function of the separation, with misalignment 0.25, $\sigma=1$.}
   \label{fig:ChernoffSincMisd}
\end{figure}

\begin{figure}[h!]  
   \centering
   \includegraphics[scale=0.38]{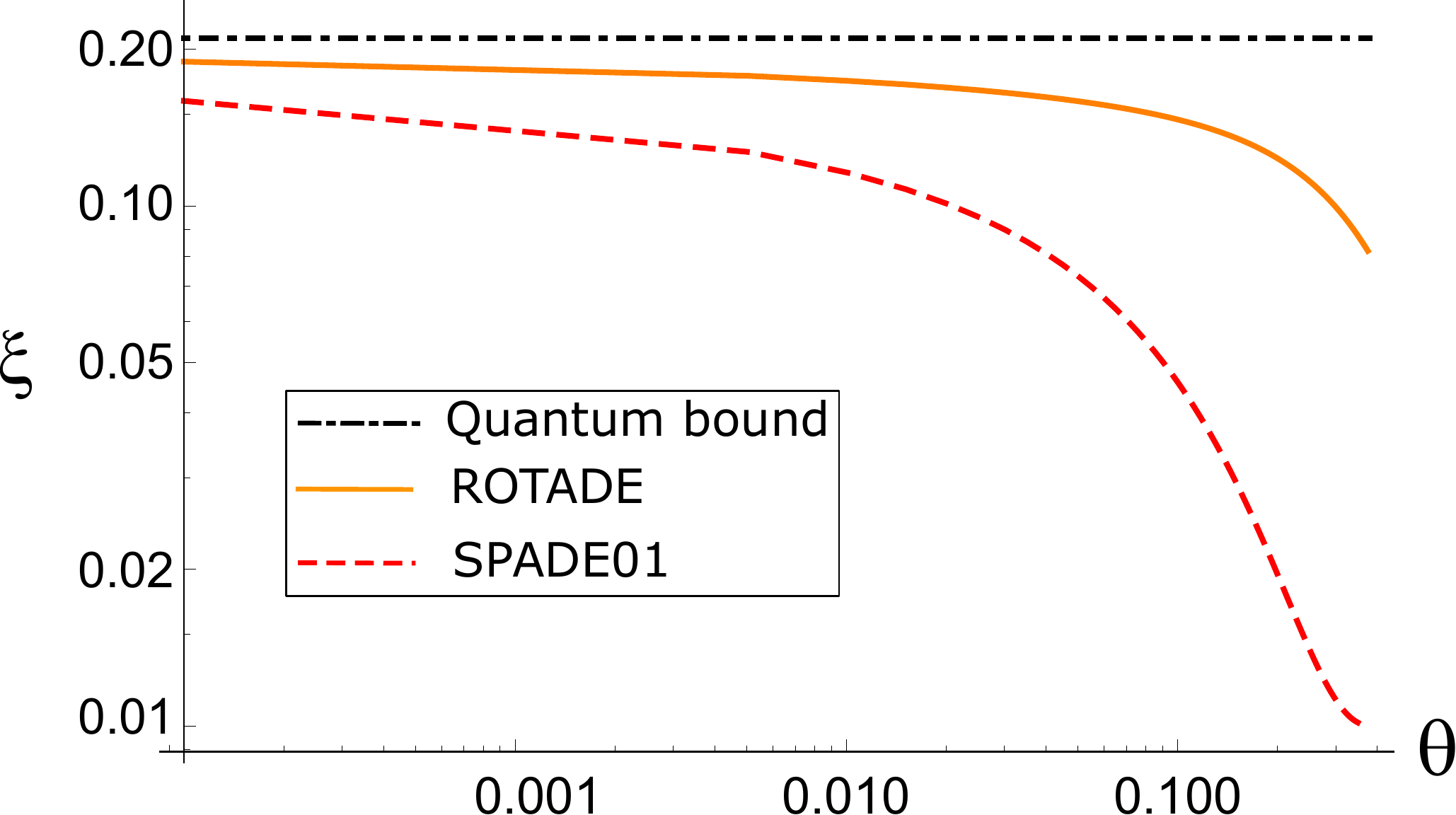}
   \caption{Chernoff exponent for Sinc PSF in function of the misalignment, for fixed separation 0.25, $\sigma=1$.}
   \label{fig:ChernoffSinMisxR}
\end{figure}

In this appendix, we present the results of discriminating one from two incoherent point sources imaged by a system with a rectangular 
aperture. The PSF of such a system is given by the Sinc function (see~\eqnref{eq:single_source_PSFb}). We compare the measurement strategies of ROTADE and SPADE$01$ with the quantum Chernoff bound in function of the misalignment, as presented in Figure~\ref{fig:ChernoffSincMisd} and the separation, in Figure~\ref{fig:ChernoffSinMisxR}. 

Similarly to the results in the main text, we verify in the limit asymptotic limit, ROTADE performs better than SPADE01.

\section{Performance of ROTADE in discrimination}
\label{appendix:Qubitperformance}
Here we analyse the performance of ROTADE for the task of discriminating one and two light sources. As ROTADE involves only thee two-dimensional subspace spanned by the zeroth and first HG modes, an intrinsic error probability arises when the incoming radiation couples into higher HG modes. This probability is useful for defining the regime of validity of the qubit model.
\begin{figure}[h!]  
   \centering
   \includegraphics[scale=0.75]{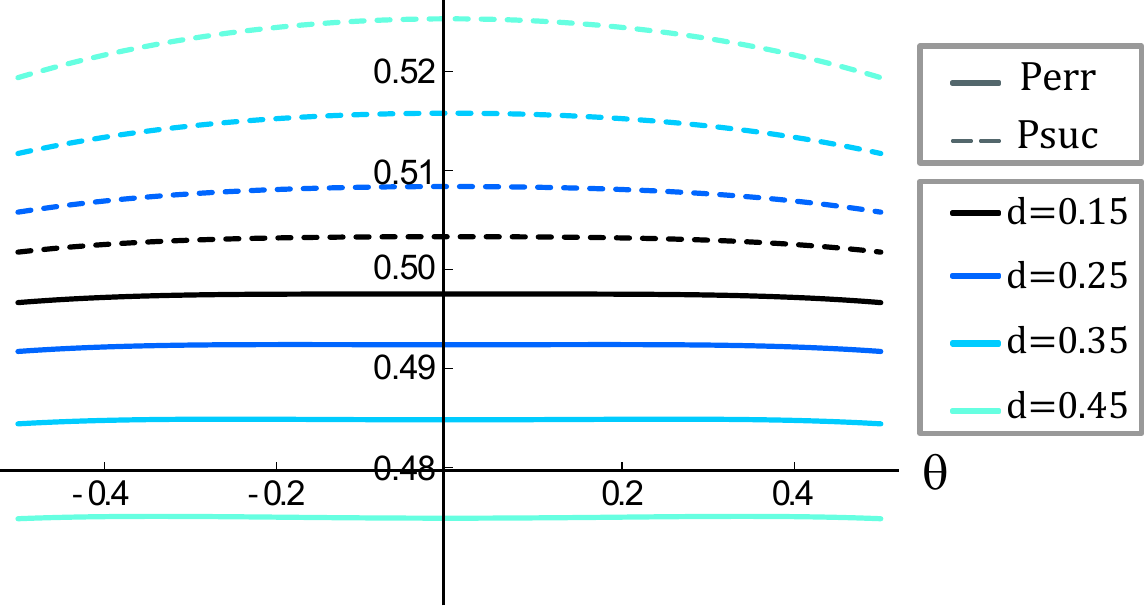}
   \caption{Error (solid) and success (dashed) probability in function of the reference position, using ROTADE for distinguishing between one and two sources, for different separations between the two sources.}
   \label{fig:PerrPsuc}
\end{figure}

For example Figure~\ref{fig:PerrPsuc} presents the error and success probabilities in the regime where the centre of each distribution are aligned $\theta=0$.  We observe that ROTADE has constant value (less than $4\%$ variation), e.g., at $\theta=0$ the error probability has value $P_{\mathrm{err}}=\frac{1}{2}\left(P_{\mathrm{err1}}+P_{\mathrm{err2}}\right)=\frac{1}{2}\left(0+e^{\frac{-d^{2}}{4\sigma^{2}}}\right)$, and the success probability $P_{\mathrm{suc}}=\frac{1}{2}\left(P_{\mathrm{suc1}}+P_{\mathrm{suc2}}\right)=\frac{1}{2}\left(1+\frac{d^{2}}{4\sigma^{2}}e^{\frac{-d^{2}}{4\sigma^{2}}}\right)$. As $d$ increases, the likelihood that photons couple to higher HG modes increases and hence the error (success) probability move further away from the priors, $0.5$. This is a consequence of the intrinsic error of the qubit model.

\begin{figure}[h!]  
   \centering
   \includegraphics[scale=0.85]{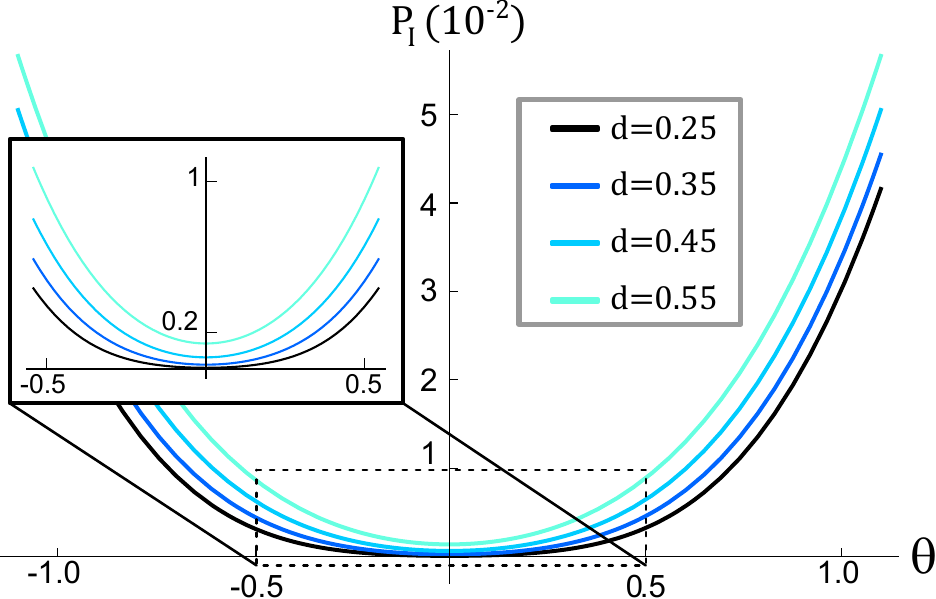}
   \caption{Intrinsic error of the qubit model ($\mathrm{P_{I}}$) in function of the misalignment $\theta$, for different separations $d$ between the sources.}
   \label{fig:PquesxR}
\end{figure}

\begin{figure}[h!]  
   \centering
   \includegraphics[scale=0.85]{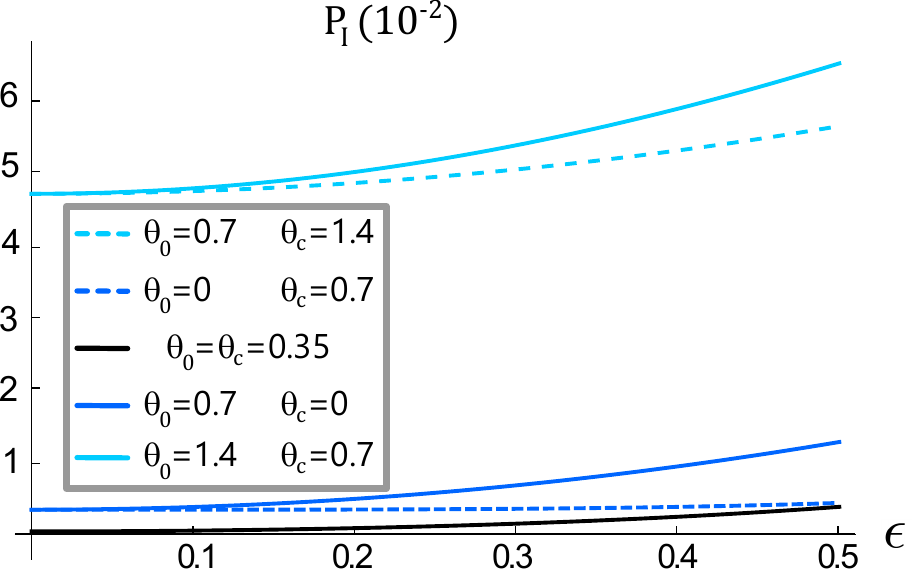}
   \caption{Intrinsic error of the qubit model ($\mathrm{P_{I}}$) in function of the separation $\epsilon$, for different values of the misalignment for the case where the center of one source is not equal with the two-source centroid ($\theta_{\mathrm{c}}\neq \theta_0$).}
   \label{fig:Pquesd}
\end{figure}

The intrinsic error is the distance between the sum of the error and success probabilities from unity. It dictates until which separation and reference position the qubit model---and consequently ROTADE---are adequate. For $\vert \theta\vert<\frac{1}{2}$, or when the separation between the sources is comparable to $\sigma$, $\epsilon<\frac{1}{2}$, this error is negligible. This features are presented in Figure~\ref{fig:PquesxR} and ~\ref{fig:Pquesd}, respectively.

In Figure~\ref{fig:PquesxR} we present the intrinsic error in function of the misalignment $\theta$. For a range of misalignments, $\vert \theta\vert<\frac{1}{2}$, ROTADE has negligible intrinsic error. Figure~\ref{fig:Pquesd} shows the intrinsic error in function of the two source separation $\epsilon$, for misaligned source distributions, i.e., the centroid of the two sources is different from the center of one source ($x_{\mathrm{c}}\neq x_0$). We observe, that the qubit model is adequate when placing the measurement in between the distribution centroids 
$x_{\mathrm{c}}\leq x_R\leq x_0$ (in between red and orange lines) and the intrinsic error of the model is minimum when $\theta_0=\theta_{\mathrm{c}}$, i.e., when the centres of the two distributions coincide.  Notice that when the centroids of the two distributions do not coincide the ROTADE measurement will, in general, depend on the separation of the two-source hypothesis.

\newpage

\bibliographystyle{apsrev4-1}
\bibliography{superres}

\end{document}